\documentclass[a4paper,fleqn,usenatbib]{mnras}
\usepackage[T1]{fontenc}
\usepackage{ae,aecompl}
\usepackage{hyperref}
\usepackage{setspace}
\usepackage{graphicx}	% Including figure files
\usepackage{amsmath}	% Advanced maths commands
\usepackage{amssymb}	% Extra maths symbols
\usepackage{xcolor}     % Peter - orange text

\newcommand{\teff}{\ensuremath{T_{\mathrm{eff}}}}
\newcommand{\logg}{\ensuremath{\log g}}

\newcommand{\lsim}{\raisebox{-1ex}{$\stackrel{{\displaystyle<}}{\sim}$}}

\newcommand{\msun}{${\mathrm{M}}_{\odot}$} 
\newcommand{\rsun}{${\mathrm{R}}_{\odot}$}

\newcommand{\mjup}{$\rm M_{\rm Jup}$}

\def\mnras{MNRAS}% Monthly Notices of the RAS
\def\aj{AJ}%
\def\apj{ApJ}%
\def\apjl{ApJ}%
\def\aap{A\&A}%
\def\pasp{PASP}%
\def\araa{ARA\&A}%
\def\apss{Ap\&SS}
\def\nat{Nature}
\title[TIC\,137608661, a new sdBV+dM reflection-effect binary]
{Filling the gap between synchronized and non-synchronized sdBs in 
short-period sdBV+dM binaries with TESS: TIC\,137608661, a new system with 
a well defined rotational splitting} 
\author[]{
Roberto Silvotti,$^{1}$\thanks{E-mail: roberto.silvotti@inaf.it}
P\'eter N\'emeth,$^{2,3}$
John~H. Telting,$^{4,5}$
Andrzej~S. Baran,$^{6,7,8}$
\newauthor
\hspace{0.0mm}
Roy~H. \O stensen,$^{7}$
Jakub Ostrowski,$^{6}$
Sumanta~K. Sahoo,$^{6,9}$
Saskia Prins$^{10,11}$
\vspace{4mm}
\\
$^{1}$INAF-Osservatorio Astrofisico di Torino, Strada dell'Osservatorio 20, 
10025 Pino Torinese, Italy\\
$^{2}$Astronomical Institute of the Czech Academy of Sciences, CZ-25165, 
Ond\v{r}ejov, Czech Republic\\
$^{3}$Astroserver.org, F\H{o} t\'er 1, 8533 Malomsok, Hungary\\
$^{4}$Nordic Optical Telescope, Rambla Jos\'e Ana Fern\'andez P\'erez 7, 38711 
Bre\~na Baja, Spain\\
$^{5}$Department of Physics and Astronomy, Aarhus University, Ny Munkegade 120,
DK-8000 Aarhus C, Denmark\\
$^{6}$ARDASTELLA Research Group, Institute of Physics, Pedagogical University 
of Krakow, ul. Podchor\k{a}\.zych 2, 30-084 Krak\'ow, Poland\\
$^{7}$Department of Physics, Astronomy and Materials Science, Missouri State 
University, 901 S. National, Springfield, MO\,65897, USA\\
$^{8}$Embry-Riddle Aeronautical University, Department of Physical Science, 
Daytona Beach, FL\,32114, USA\\
$^{9}$Nicolaus Copernicus Astronomical Centre of the Polish Academy of 
Sciences, ul. Bartycka 18, 00-716 Warsaw, Poland\\
$^{10}$Mercator Telescope, Roque de los Muchachos Observatory, 
La Palma, Spain\\
$^{11}$Instituut voor Sterrenkunde, KU Leuven, Celestijnenlaan 
200D, B-3001 Leuven, Belgium
}

\date{Accepted XXX. Received YYY; in original form ZZZ}

\pubyear{2021}

\begin{document}
\label{firstpage}
\pagerange{\pageref{firstpage}--\pageref{lastpage}}
\maketitle

\begin{abstract}
TIC\,137608661/TYC\,4544-2658-1/FBS\,0938+788 is a new 
sdBV+dM reflection-effect binary discovered by the $TESS$ space mission 
with an orbital period of 7.21 hours.
In addition to the orbital frequency and its harmonics, 
the Fourier transform of TIC\,137608661 shows many g-mode 
pulsation frequencies from the sdB star.
The amplitude spectrum is particularly simple to interpret as we immediately 
see several rotational triplets of equally spaced frequencies.
The central frequencies of these triplets are equally spaced in period with
a mean period spacing of 270.12\,s, corresponding to consecutive $l$=1 modes.
From the mean frequency spacing of 1.25\,$\mu$Hz we derive a rotation
period of 4.6 days in the deep layers of the sdB star, 
significantly longer than the orbital period.
Among the handful of sdB+dM binaries for which the sdB rotation was measured 
through asteroseismology, TIC\,137608661 is the non-synchronized system with 
both the shortest orbital period and the shortest core rotation period.
Only NY\,Vir has a shorter orbital period but it is synchronized.
From a spectroscopic follow-up of TIC\,137608661 we measure the radial 
velocities of the sdB star, determine its atmospheric parameters, 
and estimate the rotation rate at the surface of the star.
This measurement allows us to exclude synchronized rotation also in the outer 
layers and suggests a differential rotation, with the surface rotating faster 
than the core, as found in few other similar systems.
Furthermore, an analysis of the spectral energy distribution of 
TIC\,137608661, together with a comparison between sdB pulsation properties 
and asteroseismic models, gives us further elements to constrain the system.

\end{abstract}


%%%%%%%%%%%%%%%%%%%%%%%%%%%%%%% BODY OF PAPER %%%%%%%%%%%%%%%%%%%%%%%%%%%%%%%%

%\vspace{-4mm}

\section{Introduction}

Hot subdwarf B (sdB) stars are core-helium burning stars which have had 
their hydrogen-rich envelopes stripped almost completely during the
red giant phase, most likely as a result of binary interaction
\citep{Han_2002,Han_2003,2012ApJ...746..186C,Pelisoli_2020}.

Among hot subdwarfs (a class of stars that includes sdB and sdO stars, see
\citealt{Heber_2016} for a recent review), $\sim$30\% are in wide binaries 
with F/G/K companions, while $\sim$35\% are apparently 
single \citep[see e.g.][and references therein]{Silvotti_2021}.
A handful of single sdBV (=\,sdB Variable, i.e. pulsating) stars, for which 
rotation was accurately measured through asteroseismology, show typical 
rotation periods between $\sim$25 and $\sim$100 days 
\citep[and references therein]{Charpinet_2018, Reed_2018}.

The remaining fraction of hot subdwarfs ($\sim$35\%) are in 
post-common-envelope short-period binaries with M dwarf or white dwarf (WD) 
companions.
For this subclass of systems, the rotation periods from asteroseismology
appear to be shorter as the orbital period decreases \citep{Charpinet_2018}.
But only in three systems, HD\,265435, NY\,Vir and KL UMa, with orbital 
periods of only 1.65, 2.42 and 8.25 hours respectively, the sdBV primary
appears to be fully synchronized 
\citep{Pelisoli_2021,Charpinet_2008,VanGrootel+2008}, 
at least in the outer layers of the star.
At orbital periods shorter than $\sim$6 hours, a dozen of systems
fully synchronized or very close to synchronization was found with a different 
technique, measuring the sdB/sdO rotation velocity from the spectral line 
broadening (references are given in the caption of Figure~\ref{synchro}).

Theoretical calculations of tidal synchronization time-scales
fail to account for the synchronization of sdB stars 
\citep{2018MNRAS.481..715P} and, in the case of NY\,Vir,
not even a larger convective core is able to explain its synchronization 
\citep{2019MNRAS.485.2889P}.
%Following these authors, the synchronization of NY\,Vir could be explained 
%either by a partial synchronization of at least the outer layers of the star 
%already during the common envelope phase, or by higher convective mixing 
%velocities respect to those obtained with the mixing lenght theory.

The binary system described in this paper, TIC\,137608661 (alias 
TYC\,4544-2658-1 or FBS\,0938+788), is a new bright sdBV+dM binary
(Gaia EDR3 magnitude  G\,=\,11.112$\pm$0.001),
located at $\sim$256\,pc from us 
(Gaia EDR3 parallax of 3.90$\pm$0.04\,mas).
%
%(Gaia DR2 parallax of 3.95$\pm$0.08\,mas).
%
This star was discovered by the first Byurakan survey (FBS) 
and classified as hot subdwarf or white dwarf by \citet{2010MNRAS.407..681M}.
It is classified as an sdB in the Gaia DR2 catalogue of hot subluminous stars 
\citep{2019A&A...621A..38G}.
TIC\,137608661 was ``not observed to vary'' in a short time-series photometry 
run of about 10 minutes at the Nordic Telescope, with a sampling time of 8 
seconds, excluding short-period p-modes with amplitudes higher than about 1 ppt
(part per thousand, \citealt{2010A&A...513A...6O}).

In the next sections we present the results of an analysis of the $TESS$
light curve of TIC\,137608661, together with the results of a spectroscopic 
follow-up.
In section 2 the $TESS$ light curve is described and the orbital ephemeris is 
given.
In section 3 and 4 the low- and high-resolution spectroscopic observations 
are described, that allow us to measure the radial velocities 
of the sdB star, to determine its atmospheric parameters,
%effective temperature, surface gravity, and chemical abundances, 
and to estimate its surface rotational velocity.
In section 5 an analysis of the spectral energy distribution allows us to 
further characterize the sdB primary and partially also the M dwarf companion.
In section 6 a detailed analysis of the pulsational spectrum of the sdB star 
is presented, and the characteristics of the sdB star obtained from 
spectroscopy are compared with those obtained from evolutionary pulsation 
models.
In section 7 the rotation period of TIC\,137608661 is compared 
with other similar sdBs in short-period binaries that are or are not 
synchronized with their orbital period.
In section 8 we summarize our results.

%%%%%%%%%%%%%%%%%%%%%%%%%%%%%%%%%%%%%%%%%%%%%%%%%%%%%%%%%%%%%%%%%%%%%%%%%%%%%%

%\section{Constraining the binary system}

\section{$TESS$ light curve and ephemeris} 

TIC\,137608661 was observed by the {\it TESS} space mission during 
sector 14, 20 and 26, each sector being $\sim$27 days long, with a sampling 
time of 2 minutes.
We downloaded the data from the $TESS$ Asteroseismic Science Operations Center 
(TASOC)\footnote{\url{https://tasoc.dk/}} and we used the PDCSAP fluxes 
(PDC=Presearch Data Conditioning, SAP=Simple Aperture 
Photometry, see $TESS$ documentation for more details).
%
% AGGIUNGERE UNA REF SU TESS !!!!
%
After having removed some outliers and some short subsets near the sector edges
for which an instrumental trend was clearly present, 
the data we used consists of three sets with a length of 26.47, 26.32 and
24.84 days respectively, corresponding to the following epochs
(BJD$_{TBD}$--2457000): 
1683.7-1710.2 (19/07/2019-14/08/2019), 1842.5-1868.8 (25/12/2019-20/01/2020), 
and 2010.3-2035.1 (09/06/2020-04/07/2020).
When considering all three sectors together, the frequency resolution
(1.5/T, where T is the total length) is about 0.049 $\mu$Hz.

\begin{figure}
\centering
\includegraphics[width=8.5cm,angle=0]{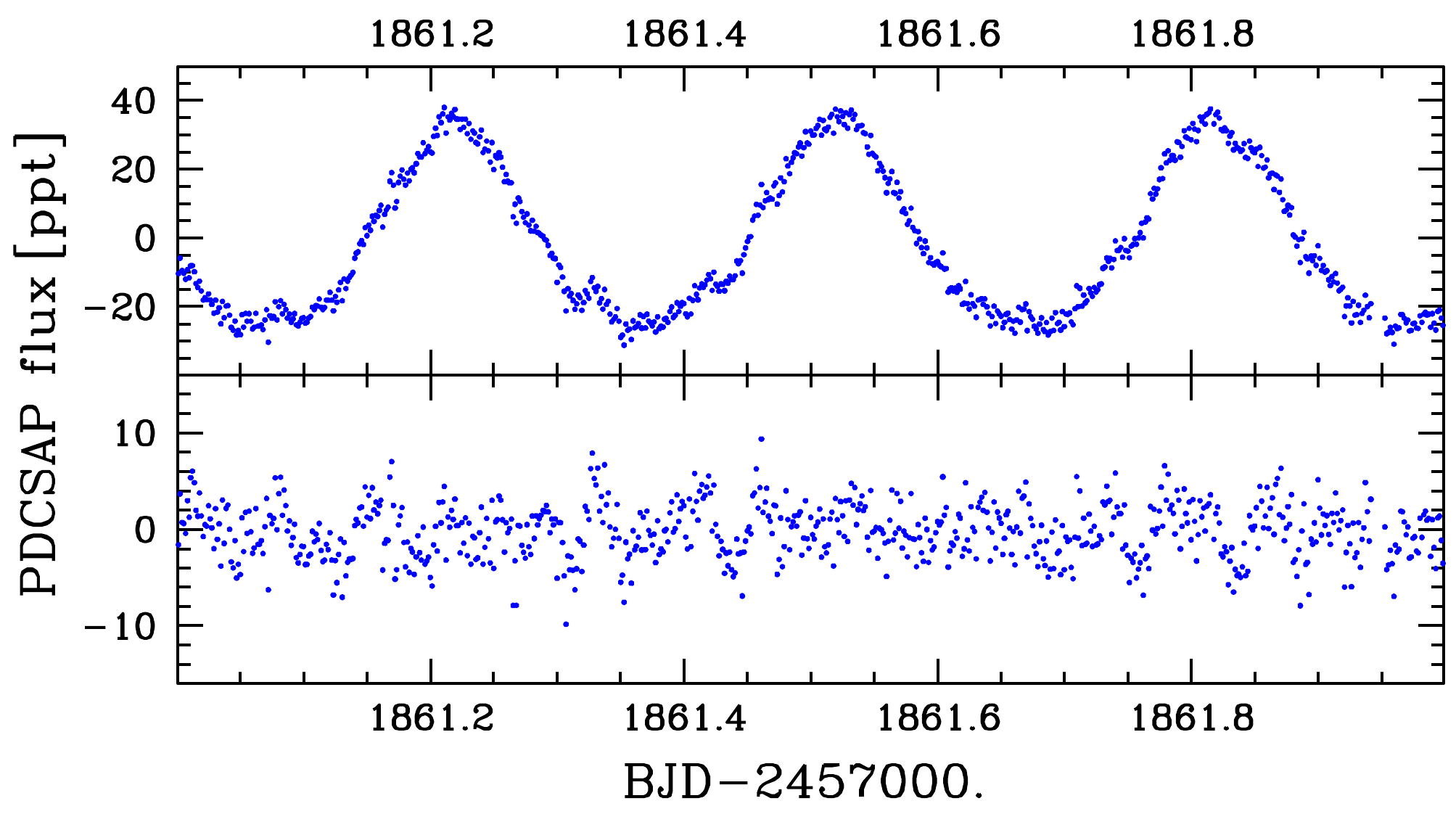}
\vspace{-3mm}
\caption{A representative 1-day section of the $TESS$ light curve. 
Top: original data. 
Bottom: residuals after removing the orbital frequency and its harmonics.}
\label{lc}
\end{figure}

A representative 1-day section of the light curve is shown in Figure~\ref{lc}.
We see that the light curve is dominated by a strong regular modulation with a 
period of 7.21 hours and a relative amplitude of 2.88\%, typical of 
a reflection effect by a cooler companion.
Moreover, when we subtract the orbital modulation (lower panel of 
Figure~\ref{lc}), we see residual low-amplitude variations suggesting that 
the sdB component is a pulsating star.

\begin{figure}
\centering
\includegraphics[width=8.5cm,angle=0]{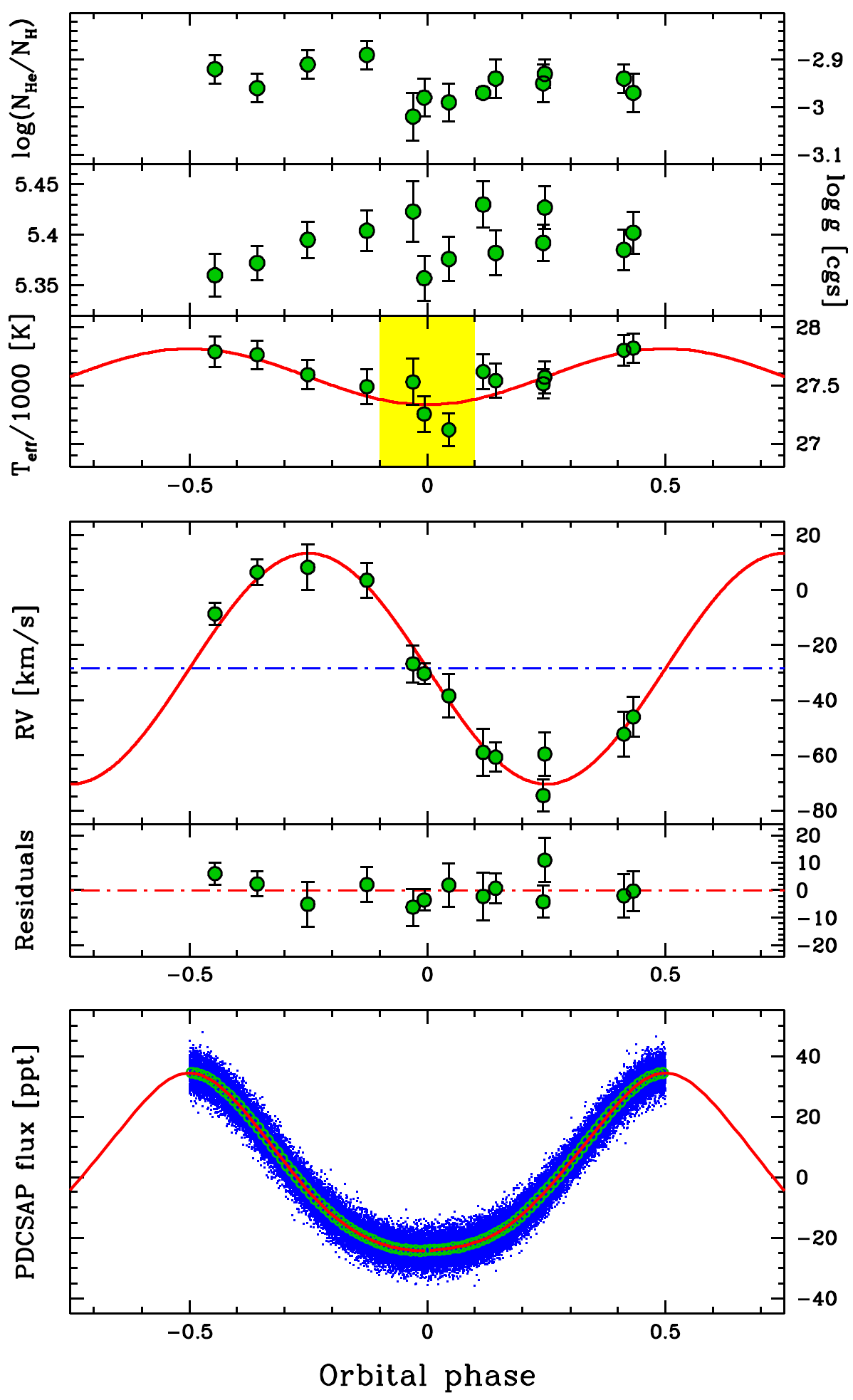}
\vspace{-3mm}
\caption{Phase resolved spectroscopy vs photometry of TIC\,137608661.
Upper panels: He abundance, \logg\ and \teff\ from LTE models 
as a function of the orbital phase. 
\teff\ shows a clear orbital modulation due to the contribution of the 
M dwarf companion and for this reason we assume as best \teff\ the mean of the 
three measurements in the phase range --0.1\,-\,0.1 (yellow rectangle), 
when the contribution of the secondary star 
%shows the hemisphere less heated by the sdB and its contribution 
is minimum.
Middle panels: radial velocities of the sdB star and residuals.
Lower panel: single $TESS$ data (blue dots), mean data in 100 phase bin
(green open dots) and best fit (red). Note that the orbital modulation is not 
perfectly sinusoidal and indeed the fit was performed using also the threee 
harmonics listed in Table~\ref{freqs}. Like in other sdB+dM systems, this 
behaviour may depend on the values of the various parameters that describe 
the so-called reflection effect
%\citep[see e.g.][for a detailed approach]{2011AJ....141...59B}.}
(see e.g. \citealt{2011AJ....141...59B} for a detailed approach).}
\label{spec_RVs_phot}
\end{figure}

The $TESS$ data were firstly used to compute the ephemeris of the system.
The following equation gives the times of the photometric maxima, when the cool
companion is behind the sdB star and shows its heated hemisphere (phase 0.5 
in Figure~\ref{spec_RVs_phot}).
BJD$_{TDB}$ 2458683.970519 corresponds to the first maximum in the $TESS$ data.

\begin{equation}
BJD_{TDB}=(2458683.970519\,\pm\,0.000035)+
\label{eq:eph}
\end{equation}

\vspace{-2mm}
\hspace{10.7mm}
(0.300420467\,$\pm$\,0.000000086)\,E

%\vspace{2mm}
%
%\noindent
%BJD$_{TDB}$=(2458683.970519\,$\pm$\,0.000035)+   % 1st maximum !!!
%
%\hspace{7.8mm}
%(0.300420467\,$\pm$\,0.000000086)\,E

%\vspace{1mm}

\section{Low-resolution spectroscopy: Radial Velocities and LTE vs non-LTE sdB atmospheric 
parameters}

\begin{figure}
\centering
\includegraphics[width=8.5cm,angle=0]{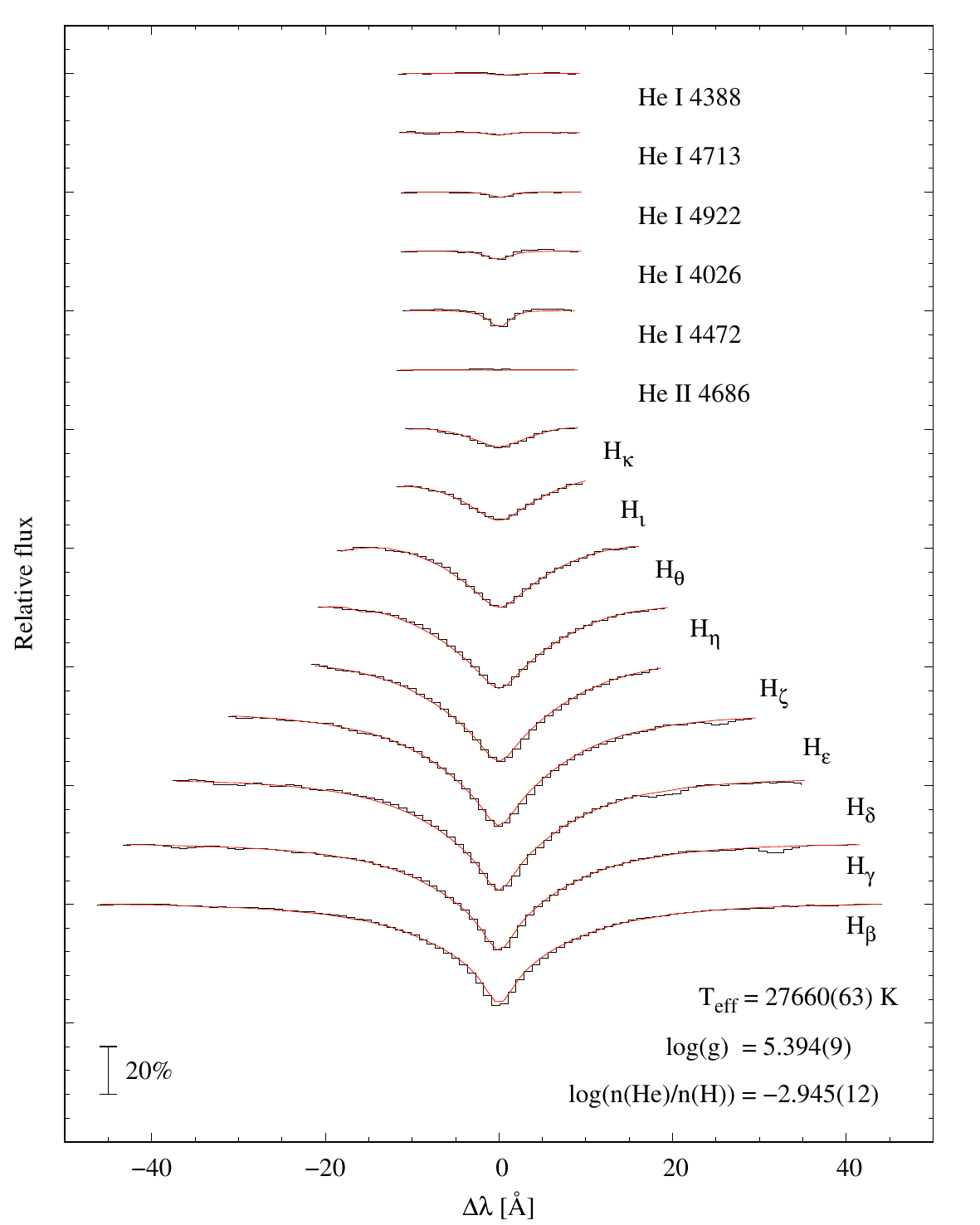}
\vspace{-3mm}
\caption{LTE fit to the mean orbit-corrected spectrum obtained from all 
the thirteen ALFOSC spectra.}
\label{LTE_spec_fit}
\end{figure}

\begin{figure*}
\centering
\includegraphics[width=\textwidth,angle=0]{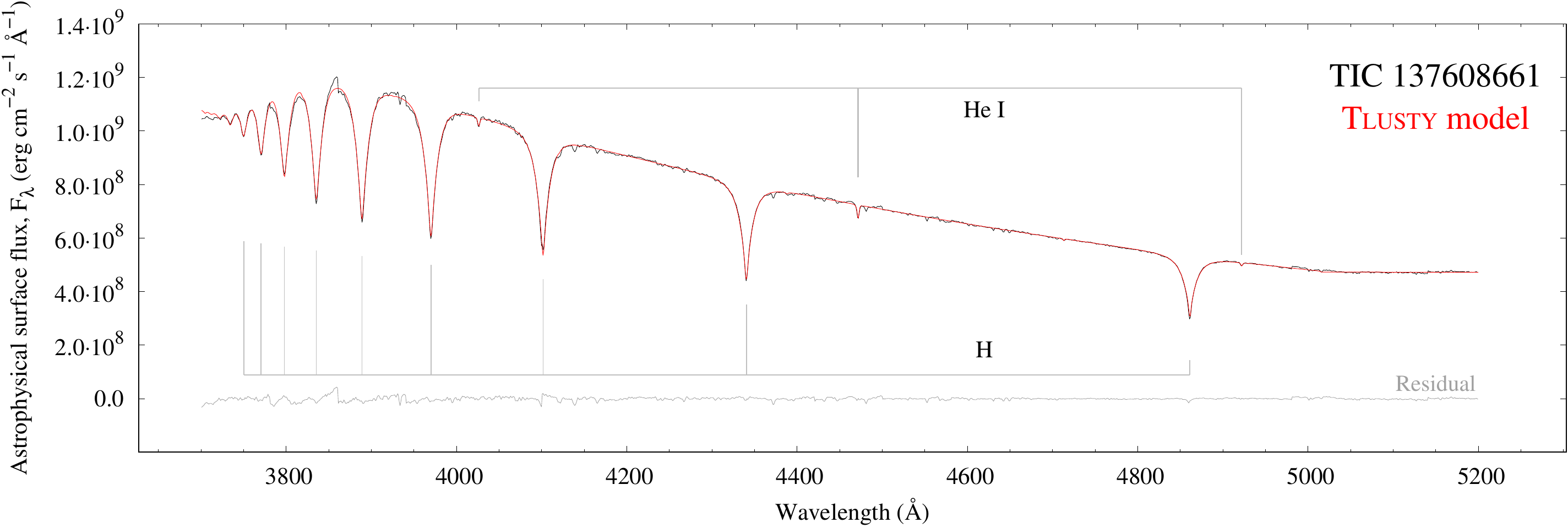}
\vspace{-3mm}
\caption{
Best-fitting {\sc Tlusty/XTgrid} non-LTE model (red) to the mean 
orbit-corrected spectrum obtained from all the thirteen ALFOSC spectra (black).
The observed spectrum has been adjusted to the continuum of the final 
theoretical model. The vertical scale corresponds to the theoretical stellar 
surface flux based on the model.}
\label{NLTE_spec_fit}
\end{figure*}

Thirteen low-resolution (R$\simeq$2000) spectra of TIC\,137608661
were obtained at the Nordic Optical Telescope (NOT, La Palma) using ALFOSC, 
250\,s exposure times, grism\#18, 0.5 arcsec slit, and CCD\#14, giving 
a resolution of 2.2 \AA\ and an approximate wavelength range 
345-535 nm.
The spectra were homogeneously reduced and analysed. 
Standard reduction steps within IRAF include bias subtraction, removal of 
pixel-to-pixel sensitivity variations, optimal spectral extraction,
and wavelength calibration based on helium arc-lamp spectra.
The peak signal-to-noise ratio of the individual spectra ranges from 80 to 250.

The spectra were taken at different orbital phases and the radial velocities
(RVs) were measured using the lines H$\beta$, H$\gamma$, H$\delta$, H8 and H9 
through a cross-correlation analysis in which we used as a template a synthetic
fit to an orbit-corrected average (all spectra were shifted to zero velocity
before averaging).
Since the RVs obtained from the last six spectra, all taken on the same night, 
showed a positive offset respect to the previous measurements, we applied 
to them a correction of --19.3 km/s.
This number was obtained by minimizing the residuals of a least-squares fit.
Then we computed a Fourier transform of the RVs, we selected the 
highest peak, and we optimized the period with a least-squares fit, obtaining 
an orbital period of 0.30058$\pm$0.00016~d, in good agreement with the 
photometric orbital period given in equation~\ref{eq:eph} (previous section).
Since the photometric period is much more precise, we then used the latter to 
optimize RV amplitude and system velocity.
We obtain a RV amplitude $K$=41.9$\pm$1.3 km/s and a system velocity 
$v$=--28.6$\pm$1.2 km/s.\footnote{Without applying any correction to the last 
six RVs, the RV amplitude does not change significantly (we obtain 44.1$\pm$3.7
km/s), while the system velocity is reduced to --20.0$\pm$2.9 km/s.}
The radial velocity fit is shown in the central panels of 
Figure~\ref{spec_RVs_phot}.

After determining the orbital RV amplitude, we shifted the spectra to the
system frame of rest, and computed a mean spectrum using all the thirteen 
spectra.
The mean spectrum reaches a peak S/N of $\sim$540 in the region 
4725\,--\,4785\,\AA.
We then used this mean spectrum to determine the physical parameters of the
target through spectroscopic model fitting.

We first did a fit using the same H/He LTE grid of \citet{Heber_2000} for
consistency with earlier studies.
We used all the Balmer lines from $H\beta$ to $H\kappa$, as well as
the five strongest \ion{He}{i} lines for the fit.
The LTE fit resulted in values of 
\teff\,=\,$27660\pm$64\,K, \logg\,=\,$5.394\pm0.009$, 
log${\rm (N_{He}/N_{H})}$\,=\,$-2.945\pm0.012$.
The errors listed on the measurements are the formal errors of the fit, which 
reflect only the signal-to-noise ratio of the mean, and not
any systematic effects caused by the assumptions underlying those models.
The best fit of the mean spectrum is shown in Figure~\ref{LTE_spec_fit}.

Then, as a next step, we performed a fit of each individual spectrum in
order to measure the variations of the atmospheric parameters 
as a function of the orbital phase, which are shown in 
Figure~\ref{spec_RVs_phot}, together with RVs and $TESS$ photometric data.
Since \teff\ shows a clear orbital modulation due to the contribution 
of the M dwarf companion, we assume as best \teff\ the mean of the three 
measurements near phase 0, when the secondary star stands in front of the
sdB primary and its contribution is minimum.
Instead, for \logg\ and log${\rm (N_{He}/N_{H})}$, we use the mean of all 13 
measurements. 
Our best values for the sdB atmospheric parameters are:
\teff=$27300\pm$200\,K, \logg=$5.39\pm$0.04, 
log${\rm (N_{He}/N_{H})}$=$-2.95\pm0.05$.

To determine the atmospheric parameters of TIC\,137608661 in non-Local 
Thermodynamic Equilibrium (non-LTE), we fitted the co-added ALFOSC 
spectrum with synthetic spectra calculated from {\sc Tlusty} models (v207; 
\citealt{Hubeny_2017, Lanz_2007}). 
The models include opacities from H, He, C, N, O, Ne, Mg, Si, and Fe. 
The spectral analysis was done with a steepest-descent spectral analysis 
procedure, implemented in the {\sc XTgrid} code \citep{Nemeth_2012}. 
The procedure is a global fitting method that simultaneously 
reproduces all line profiles with a single atmosphere model.
{\sc XTgrid} calculates new model atmospheres and corresponding synthetic 
spectra iteratively in the direction of decreasing chi-squares. 
The synthetic spectra are normalized in 80\,\AA\ sections to the observation 
to reduce the effects of the uncalibrated continuum flux on the parameter 
inference. 
Figure \ref{NLTE_spec_fit} shows the best-fit non-LTE model to the mean 
ALFOSC spectrum.
The best-fit is obtained when the relative changes of all model 
parameters decrease below 0.5\%.
Next, error calculations are performed: while for He abundances
the error bars are evaluated in one dimension, for $T_{\rm eff}$ and $\log{g}$
error calculations are performed by mapping the chi-square surface around 
the best fit as in Figure~\ref{errors}.
%as in Figure~\ref{errors}.}
%Figure~\ref{correlation}.}

%Next, error calculations are performed by mapping the chi-square surface 
%around the best-fit. 
%For abundances, the error bars are evaluated in one dimension, and parameter 
%correlations are included for $T_{\rm eff} $ and $\log{g}$ as in 
%Figure~\ref{correlation}.

The spectroscopic parameters obtained from LTE and non-LTE models 
are summarized in Table \ref{tab:summary}.
The error bars are statistical. 
Systematic errors can be estimated from the differences between the two 
independent analyses. 
While the surface gravity and helium abundance agree within error bars, the 
effective temperature is slightly higher in the non-LTE analysis.
This difference may partially arise from the fact that in the LTE 
analysis we used only three spectra close to phase zero, while in non-LTE we 
used all the spectra since the orbital modulation of \teff\ could not be 
measured, most likely because the global fitting procedure 
smears out these effects.

%The orbital modulation of the parameters observed in LTE could not be measured
%in non-LTE. Most likely the global fitting procedure smears out these 
%effects.

\begin{table}
\setstretch{1.1}
\centering
\caption{Spectroscopic parameters, LTE vs non-LTE models.     
\label{tab:summary}}
\begin{tabular}{l|rr}
\hline
Parameter & LTE~~~~ & non-LTE~~~\\
\hline
$T_{\rm eff}$ (K)          & ~~~27300$\pm200^1$ & ~~27960$\pm100^2$\\
$\log{g}$  (cm s$^{-2}$)   & $5.39\pm0.04$      & $5.42\pm0.04$\\
$\log{n{\rm He}/n{\rm H}}$ & $-2.95\pm0.05$     & $-2.89\pm0.05$\\
%$\log{n{\rm C}/n{\rm H}}$  &                & $-5.68\pm0.25$ \\
%$\log{n{\rm N}/n{\rm H}}$  &                & $-4.53\pm0.16$ \\
%$\log{n{\rm O}/n{\rm H}}$  &                & $-4.36\pm0.03$ \\ 
%$\log{n{\rm Ne}/n{\rm H}}$ &                & $<-3.6$        \\ 
%$\log{n{\rm Mg}/n{\rm H}}$ &                & $-4.83\pm0.17$ \\ 
%$\log{n{\rm Si}/n{\rm H}}$ &                & $-5.93\pm0.15$ \\
%$\log{n{\rm Fe}/n{\rm H}}$ &                & $-4.73\pm0.22$ \\
%%%%%%%%%%%%%%%%%%%%%%%%%%%%%%%%%%%%%%%%%%%%%%%%%%%%%
%$T_{\rm eff}$ (K)          & $27300\pm200$  & $27960^{+10}_{-80}$   \\
%$\log{g}$  (cm s$^{-2}$)   & $5.39\pm0.04$  & $5.436^{+0.022}_{-0.054}$  \\
%$\log{n{\rm He}/n{\rm H}}$ & $-2.95\pm0.05$ & $-2.910^{+0.073}_{-0.033}$ \\
%$\log{n{\rm C}/n{\rm H}}$  &                & $-5.481^{+0.050}_{-0.455}$ \\
%$\log{n{\rm N}/n{\rm H}}$  &                & $-4.590^{+0.222}_{-0.104}$ \\
%$\log{n{\rm O}/n{\rm H}}$  &                & $-4.342^{+0.012}_{-0.047}$ \\ 
%$\log{n{\rm Ne}/n{\rm H}}$ &                & $<-3.6$        \\ 
%$\log{n{\rm Mg}/n{\rm H}}$ &                & $-4.827^{+0.158}_{-0.168}$ \\ 
%$\log{n{\rm Si}/n{\rm H}}$ &                & $-5.831^{+0.049}_{-0.244}$ \\
%$\log{n{\rm Fe}/n{\rm H}}$ &                & $-4.814^{+0.306}_{-0.128}$ \\
\hline
\multicolumn{2}{l}{$^1$~From only 3 spectra near orbital phase 0.}\\
\multicolumn{2}{l}{$^2$~From all 13 spectra. See text for more details.}
\end{tabular}
\end{table}

\begin{figure*}
\centering
\includegraphics[width=\textwidth]{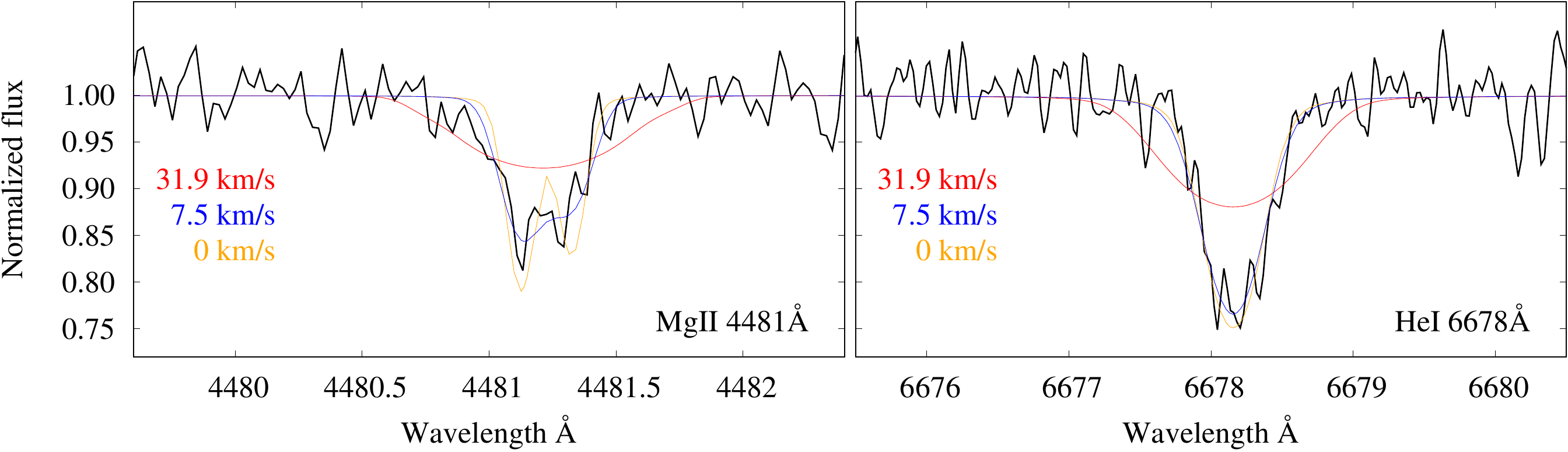}
\vskip 5pt
\includegraphics[width=\textwidth]{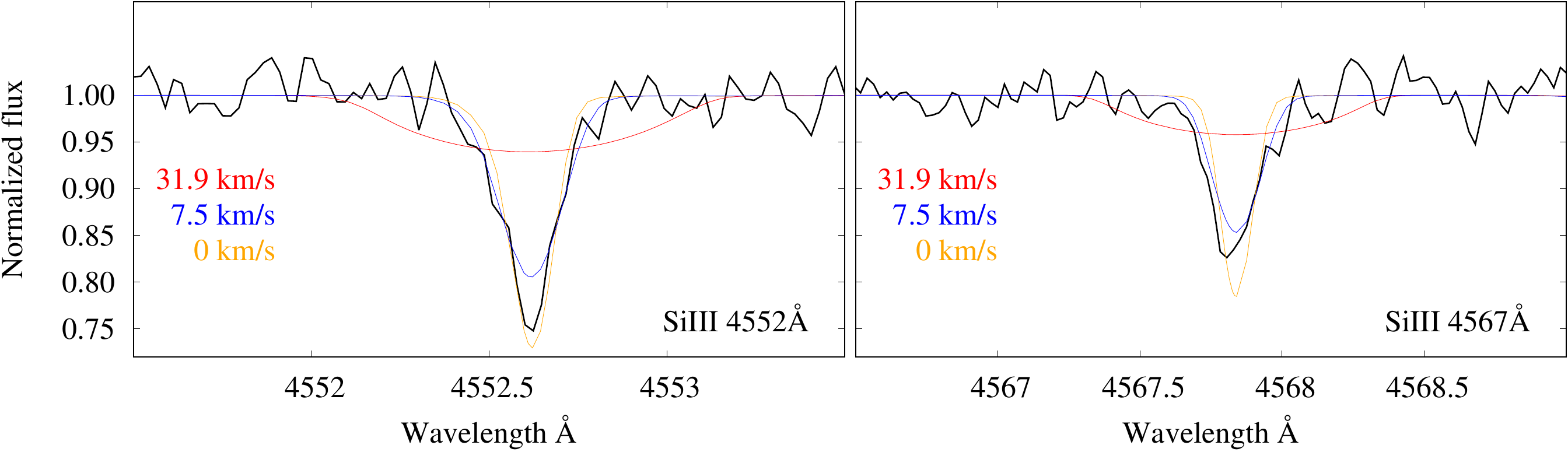}
\caption{
Line fits to the Mg\,{\sc ii} 4481 \AA, He\,{\sc i} 
6678 \AA, Si\,{\sc iii} 4552 \AA\ and Si\,{\sc iii} 4567 \AA\ lines with 
different projected rotation velocities. A synchronous rotation with the binary
orbit at $V_{\rm e}\sin{i}=31.9$ km/s 
(assuming $R$\,=\,0.209 R$_\odot$ and an inclination of 65$\degr$, see section 
5 and 6.2) can be ruled out.
A rigid stellar rotation at $\sim$2.1 km/s (from P$_{\rm rot}$=4.6~d, see
section 6.2) is not represented since the line profiles would be hardly  
distinguishable from the zero-rotation profiles.
The fit for the entire HERMES spectrum is shown in Figure~\ref{HR_fit}.
%is available online at: \url{https://astroserver.org/UX87S2}.
}
\label{fig:line}
\end{figure*}

\begin{figure}
\centering
\includegraphics[width=8.5cm,angle=0]{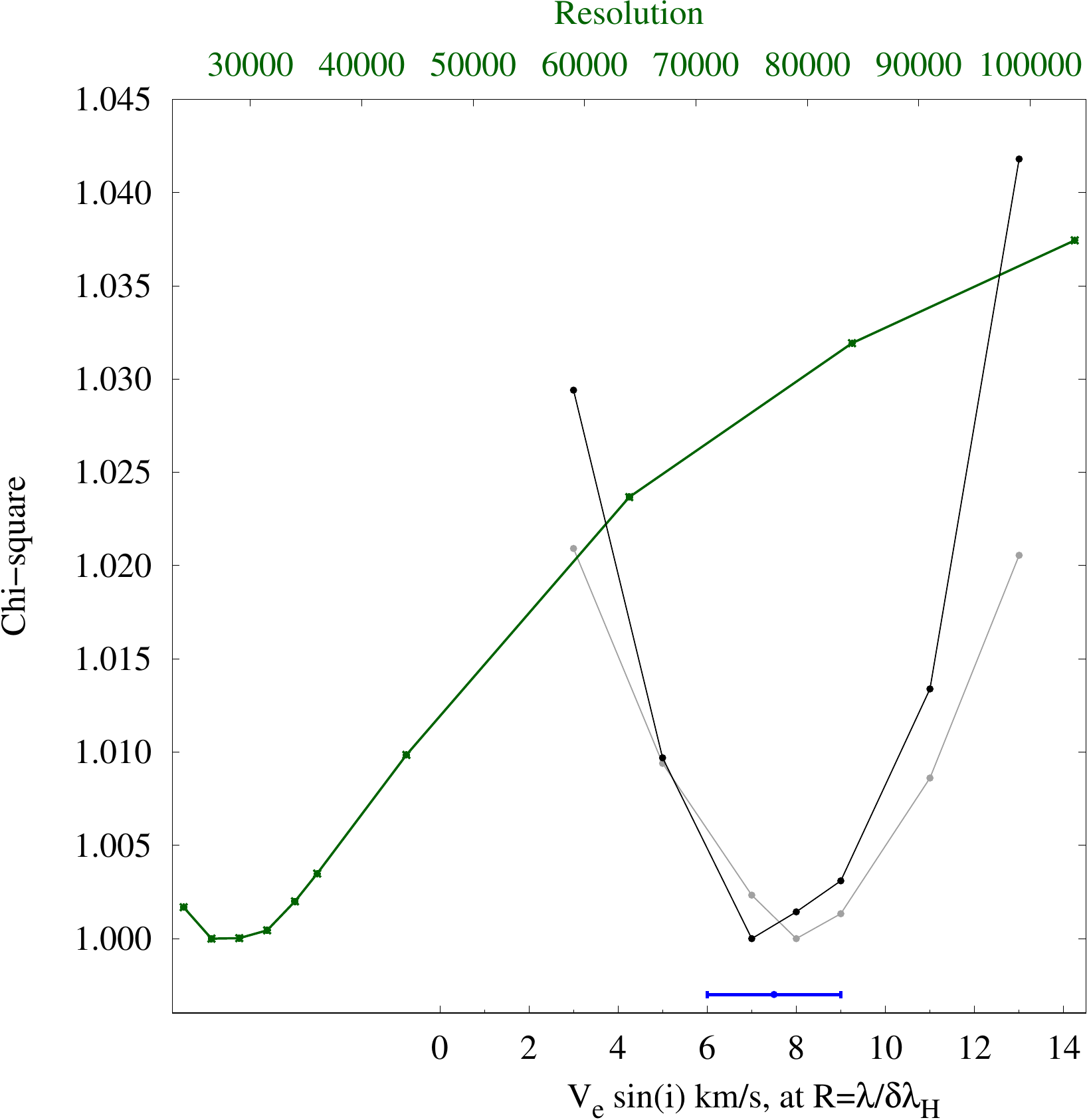}
\caption{
Chi-square values with respect to the projected equatorial 
rotation velocity (bottom axis). The black curve is based on the lines in 
Figure~\ref{fig:line} and the grey curve is for all selected regions 
containing metal line in the HERMES spectrum 
(see Table~\ref{WLranges}). 
Both curves show the same minimum, where the blue error bar represents our 
adopted final value. 
The green curve shows chi-square values at different resolutions (top axis)
and zero rotation. 
It confirms that there is an extra broadening in the spectrum and the line 
profiles cannot be reproduced by instrumental broadening alone. The observed 
broadening due to projected rotation corresponds to an instrumental profile of 
$R\approx30\,000$.
}
\label{fig:vsini}
\end{figure}

%\begin{figure}
%\centering
%\includegraphics[width=\linewidth,angle=0]{Teff_logg.png}
%%\vspace{-5mm}
%\caption{ 
%Non-LTE $T_{\rm eff}-\log{g}$ correlation and two dimensional error 
%determination for TIC\,137608661. 
%The color bar shows the chi-square variations with the parameters. 
%The contours are for 60, 90 and 99\% confidence intervals and the white error 
%bars represent the adopted final errors.}
%\label{correlation}
%\end{figure}

\section{High-resolution spectroscopy: sdB surface rotation and metal 
abundances}

In order to complement our seismic internal rotation determination
described in section 6.2, and try to measure (or at least to put an upper 
limit to) the rotation rate at the surface of the star from the 
rotational line-broadening $V_{\rm e}\sin{i}$ of TIC\,137608661, we obtained 64
high-resolution spectra, R=85000, using the HERMES instrument 
\citep{2011A&A...526A..69R} at the Mercator telescope.
The spectra were obtained from 2021-05-07 to 2021-06-06 with an exposure time
of 600\,s.
We used the wavelength- and barycentric corrected, cosmic-ray clipped, 
order-merged HERMES-pipeline product\footnote{see 
\url{http://mercator.iac.es/instruments/hermes/drs/} for more details.}.

As the orbital radial velocity varies during the exposures, only
observations near the RV minima or maxima at orbital phases 0.25 and 0.75 
(see Figure \ref{spec_RVs_phot}) have minimal orbital broadening, which is 
essential for our attempt to measure $V_{\rm e}\sin{i}$. 
Therefore we used our TESS ephemeris (equation \ref{eq:eph}) to predict the 
best times to acquire the spectra.
We obtained 18 spectra close to the orbital phase of RV minimum, and
15 spectra close to the orbital phase of RV maximum. 
We co-added these 33 spectra after shifting them to remove the orbital RV 
variation, and we had to apply K=42.26 km/s and a system velocity of -28.41 
km/s in order to minimise the velocity difference between the co-added 
orbit-corrected spectra. 
These values are in very good agreement with the orbital solution found in
section 3 from the ALFOSC@NOT spectra.

The 33 selected spectra have S/N between 4.5 and 11.7; their orbital phase 
at mid-exposure falls in the ranges 0.2162--0.2806 or 0.7147--0.7849 
(using our TESS ephemeris), and the RV variation during the exposures, 
according to our orbital solution, is less than 1.35 km/s, which is much 
smaller than the instrumental broadening of FWHM$\sim$3.5\,km/s.
After co-adding these well-selected 33 orbit-corrected spectra, we reach a 
continuum peak-S/N around 50. The useful wavelength range is 3900--8950\,\AA, 
while the bluest and reddest discernible sharp features are CaII 
$\lambda$3934\,\AA\ and HeI $\lambda$7065\,\AA.

We fitted the co-added HERMES spectrum using {\sc XTgrid} for abundances and 
$V_{\rm e}\sin{i}$.  
For the wavelength dependent dispersion of the spectrograph we used 
$\delta\lambda_H=0.0087+9.762\times10^{-6}\,\lambda$, as empirically
derived from the width of the ThAr lines in an extrated ThAr spectrum.

Figure~\ref{fig:line} shows fits to selected metal lines at two different 
projected rotation velocities and with zero rotation for reference. 
The fit for the entire HERMES spectrum can be seen in Figure~\ref{HR_fit}.
%also available online\footnote{\url{https://astroserver.org/UX87S2}}.
Our models show that a rotation synchronized with the orbit, corresponding to 
31.9 km/s (assuming $R$\,=\,0.209~R$_\odot$ and $i$=65$\degr$, 
see section 5 and 6.2), can be ruled out. 
However, we find that our models best describe the data for 
$V_{\rm e}\sin{i}$\,$\simeq$\,7.5 km/s. 
Due to variations in the S/N ratio and continuum placement, the individual 
lines in Figure~\ref{fig:line} do not reflect the projected rotation well. 
Therefore we selected regions of the spectrum containing sharp metal lines, 
that are the most sensitive for rotational broadening.
These regions are listed in Table~\ref{WLranges}.
Repeating the analysis for these regions confirm a clear minimum in chi-square
corresponding to $V_{\rm e}\sin{i}=7.5\pm1.5$ km/s,
as shown in Figure~\ref{fig:vsini}.
Considering that part of the line broadening may be caused also by other 
phenomena like instrumental broadening, orbital smearing, micro and macro 
turbulence, this value should be considered as an upper limit.
Among these phenomena, macro turbulence caused by pulsations should 
be a minor effect.
Indeed, g-mode pulsations in sdB stars produce typical RV variations of less 
than 1 km/s \citep{Silvotti_2020}.
In HD\,4539, that has pulsation amplitudes of the main peaks very similar to
TIC\,137608661, the RV variations do not exceed 200-300 m/s 
\citep{Silvotti_2019}.
Also microturbulence is expected to be small in sdB stars: it was constrained 
to $\lesssim$ 2 km/s in HD\,188112 \citep[although HD\,188112 is a peculiar
low-mass sdB, quite different from TIC\,137608661]{2016A&A...585A.115L}.
In conclusion our analysis suggests a differential rotation for 
TIC\,137608661, with the envelope rotating faster than the core at a
projected rotation velocity not higher than 7.5 km/s.
A rigid rotation of the star, which would imply a projected rotation velocity 
of $\sim$2.1 km/s (from $R$=\,0.209 R$_\odot$, P$_{\rm rot}$=4.6~d, 
$i$=65$\degr$, see section 5 and 6.2) appears unlikely but can not
be completely excluded.
%
%\footnote{\color{red} Macro turbulence caused by pulsations should 
%be a minor effect, at least an order of magnitude smaller. 
%For example, in HD\,4539, with very similar amplitudes of the main pulsation 
%peaks, the RV variations do not exceed 200-300 m/s \citep{Silvotti_2019}}),
%this value should be considered as an upper limit.
%Therefore, even if this result suggests a differential rotation for 
%TIC\,137608661, we cannot completely rule out a rigid rotation, 
%which would imply a projected rotation velocity of $\sim$2.1 km/s 
%(from $R$=\,0.209 R$_\odot$, P$_{\rm rot}$=4.6~d, $i$=65$\degr$, 
%see section 5 and 6.2).
%
%{\bf Even though we optimised the observations for obtaining a conclusive
%value for $V_{\rm e}\sin(i)$, we find from our spectral model fits
%that the rotational line broadening is not outstanding significantly
%with respect to other sources, intrinsic and observational, of line
%broadening.
%}

The fitting procedure includes all lines from the {\sc Synspec} line list
that show up in the spectrum, which contribute to the fit based on the 
strength of each line and the S/N ratio of the observation at that wavelength. 
Table~\ref{tab:abn} lists metal abundances derived from the HERMES spectrum. 
The quoted errors are statistical only, calculated in one dimension by mapping 
the chi-square around the best fit.
The measured metal abundances agree with the observed abundance pattern in
sdB stars, for which light metals are typically sub-solar while iron is near 
the solar value \citep{2013A&A...549A.110G}.

\begin{table}
\setstretch{1.05}
\centering
\caption{Surface metal abundances from high-resolution HERMES spectra and 
non-LTE models. Solar abundance fractions are given with respect to 
\citep{2009ARA&A..47..481A}. 
\label{tab:abn}}
\begin{tabular}{l|cc}
\hline
Element & $\log{n{\rm X}/n{\rm H}}$ & $\log{\epsilon/\epsilon_\odot}$\\
\hline
C  & $-4.44\pm0.16$ & -0.74\\
N  & $-4.49\pm0.03$ & -0.29\\
O  & $-4.44\pm0.08$ & -1.20\\ 
Ne & $-4.29\pm0.25$ & -0.28\\ 
Mg & $-5.23\pm0.11$ & -0.72\\ 
Si & $-5.54\pm0.12$ & -1.16\\
Fe & $-4.30\pm0.20$ & \ 0.39\\
\hline
\end{tabular}
\end{table}

\section{Spectral Energy Distribution}

\begin{figure*}
\centering
\includegraphics[width=1.008\textwidth,angle=0]{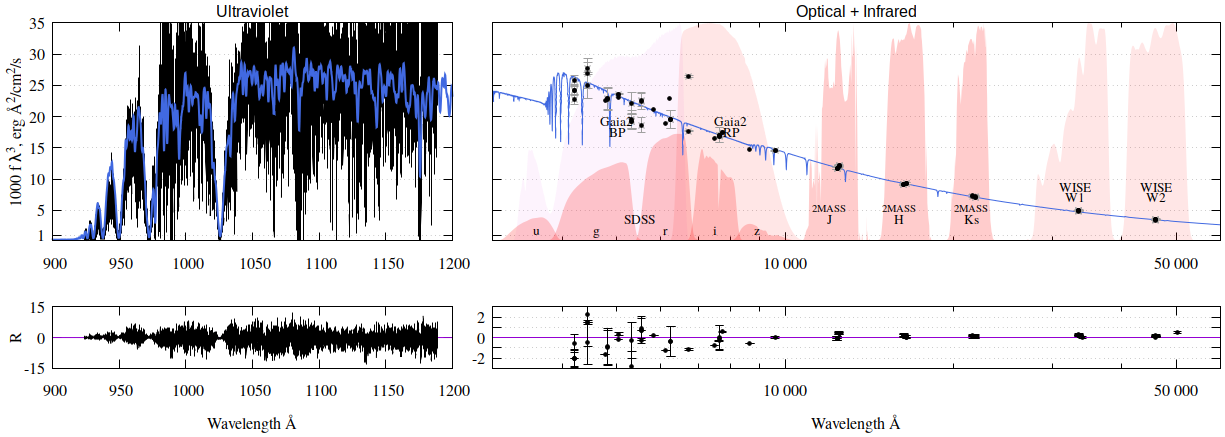}
\vspace{-5mm}
\caption{
Spectral energy distribution of TIC\,137608661 from the far-ultraviolet (FUSE 
spectrum, left) to the optical and infrared (WISE, right).
All photometric data points were taken from the VizieR data service and 
dereddened for $E(B-V)=0.017$ mag.
The shaded regions mark various filter pass-bands.
The figure shows the qualitative match between our final spectral model and 
broadband photometry. 
The synthetic spectrum was normalized to the observations in the WISE W1 band 
(33500\AA).
The SED shows that the FUSE spectrum along with the photometric measurements 
are consistent with a single stellar component.
}
\label{sed}
\end{figure*}

Spectral energy distributions (SED) provide a way to evaluate the 
contributions of binary members to the observed flux.
Hot subdwarf stars with F- or G-type companions can be described with two 
components, which contribute nearly equally to the observed flux in the 
optical.
Late K- and M-type companions remain nearly invisible next to a hot subdwarf.
The only exceptions are those in close orbits, when the irradiation by 
the hot subdwarf is able to form a hot spot on the companion.
The strength of irradiation depends on the radii of the components, their 
separation, as well as the temperature and the spectral properties of the 
irradiating star (see e.g. \citealt{1993ApJ...419..344M} for a simple 
analytical model).
In TIC\,137608661 the amplitude of the reflection effect is relatively low,
only $\sim$6\% (57.5 ppt)
%
% 5\% (50 ppt)
%
when we consider the total amplitude, i.e. the difference between maximum 
and minimum flux.
In HW\,Vir and NY\,Vir this contribution is about 20\% in the optical, 
but a precise spectral characterization of the cool companion is still 
difficult, while in the sdO+dM binary AA\,Dor it was possible 
\citep{Vuckovic_2016} despite the reflection effect amplitude is only 
$\sim$7\%.

For a SED analysis {\sc XTgrid} collects photometric data from the VizieR 
Photometry Viewer\footnote{\url{http://vizier.u-strasbg.fr/vizier/sed/}} 
around 2 arcseconds of the target and shifts the synthetic SED calculated from 
the spectral modeling to the photometric data.
The interstellar extinction toward TIC\,137608661 is low, 
$E$\,($B$--$V$)\,=\,0.017 mag \citep{Schlafly_2011}.
Figure \ref{sed} shows that the match over the optical and infrared regions 
is excellent and the SED can be modeled with a single hot component.
However, a cool ($T_{\rm eff}\lesssim4000$\,K) M dwarf can remain invisible 
in the SED.
The ultraviolet region is very important because the SED of the subdwarf 
peaks there.
The only measurement from GALEX is significantly off, most 
likely due to the bright non-linearity of the GALEX photometry 
\citep{Morrissey_2007}. 
Fortunately, there are FUSE observations of the star 
(e.g.: FUSE Program ID: G061, PI: Pierre Chayer), which confirm the far-UV 
flux level (left panels of Figure \ref{sed}).

With an independent distance measurement provided, the SED analysis returns
radius, luminosity and mass of the hot subdwarf.
We used the Gaia EDR3 distance $d$\,=\,256.5\,$\pm$\,2.6 pc and found 
an angular diameter $\log{\theta}$\,=--10.434\,$\pm$\,0.070 rad for the sdB 
in TIC\,137608661, which results in a stellar radius 
$R$\,=\,0.209\,$\pm$\,0.005 R$_\odot$.
Then, adopting \teff=27960~K and \logg=5.42 from non-LTE results,
we obtain $L$\,=\,23.98\,$\pm$\,1.09 L$_\odot$ and
$M$\,=\,0.419\,$\pm$\,0.041 M$_\odot$.
%
%$M$\,=\,0.441\,$\pm$\,0.041 M$_\odot$, $R$\,=\,0.214\,$\pm$\,0.005 R$_\odot$, 
%{\color{red} $L$\,=\,25.34\,$\pm$\,1.09 L$_\odot$ and an angular diameter 
%of $\log{\theta}$\,=--10.434\,$\pm$\,0.070 rad for the sdB in TIC\,137608661.}

%With an independent distance measurement provided, the SED analysis returns 
%the radius, mass, and luminosity of the hot subdwarf.
%We used the Gaia EDR3 distance $d$\,=\,256.5\,$\pm$\,2.6 pc 
%{\color{red} to find} {\color{red} an angular diameter of 
%$\log{\theta}=-10.434\pm0.070$ rad}
%{\color{red} and obtain} $R$\,=\,0.214\,$\pm$\,0.005 R$_\odot$,
%$M$\,=\,0.441\,$\pm$\,0.041 M$_\odot$ and
%{\color{red} $L$\,=\,25.34\,$\pm$\,1.09 L$_\odot$ for the sdB in 
%TIC\,137608661.}

%B=10.97, V=11.34, J=11.759, K=11.946
Following \cite{Deca_2012}, the $B$--$V$\,=\,--\,0.37 and 
$J$--$K_s$\,=\,--\,0.187 mag 
color indices suggest a very low contribution from the companion,
indicating that it must be a low-mass main sequence star.
A white dwarf companion is excluded since it would not give rise to the
reflection effect that we see in TESS data, while it would produce ellipsoidal 
variations which are not detected.
Moreover, from the binary mass function, a WD companion would imply an 
inclination of the system lower than $\sim$15$\degr$, which is not compatible 
with the inclination of (65$^{+10}_{-20}$)$\degr$ found from the seismic 
analysis in section 6.2.
%\cite{Maxted_2002} have derived a relationship between the separation and 
%relative radii of the components that can be used to estimate the amplitude 
%of the reflection effect for different companion types. 
%At a similar orbit, a white dwarf would produce a reflection amplitude 
%100 times less than a cool main sequence star. 
%This factor is large enough, that even at an orbital inclination as low as 
Further spectral characterization of the companion would require high SNR 
infrared spectroscopy.

%%%%%%%%%%%%%%%%%%%%%%%%%%%%%%%%%%%%%%%%%%%%%%%%%%%%%%%%%%%%%%%%%%%%%%%%%%%%%%

\section{SdB pulsations}

\subsection{Fourier analysis}

\begin{figure*}
\centering
\includegraphics[width=1.008\textwidth,angle=0]{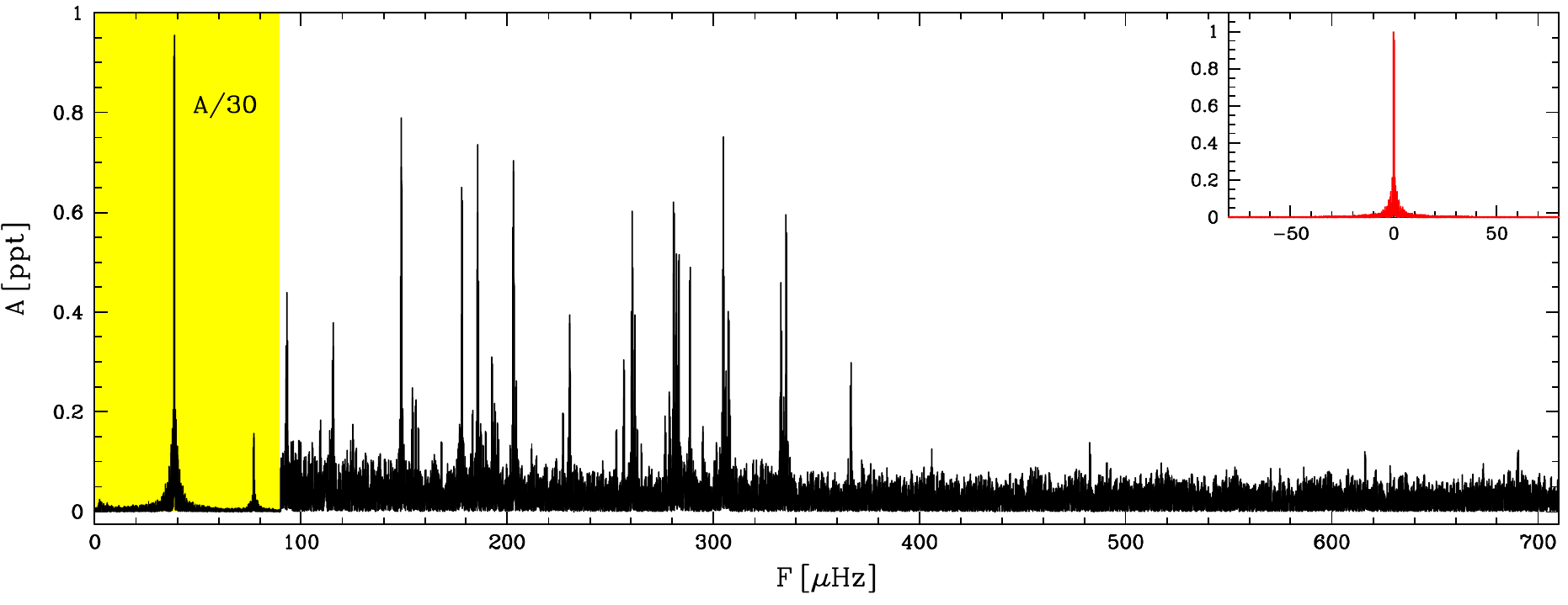}
\vspace{-5mm}
\caption{Amplitude spectrum of TIC\,137608661 using all three sectors together.
The yellow section corresponds to the orbital frequency and its first harmonic.
The upper-right inset is the window function (in red).}
\label{dft}
\end{figure*}

\begin{figure}
\centering
\includegraphics[width=0.48\textwidth,angle=0]{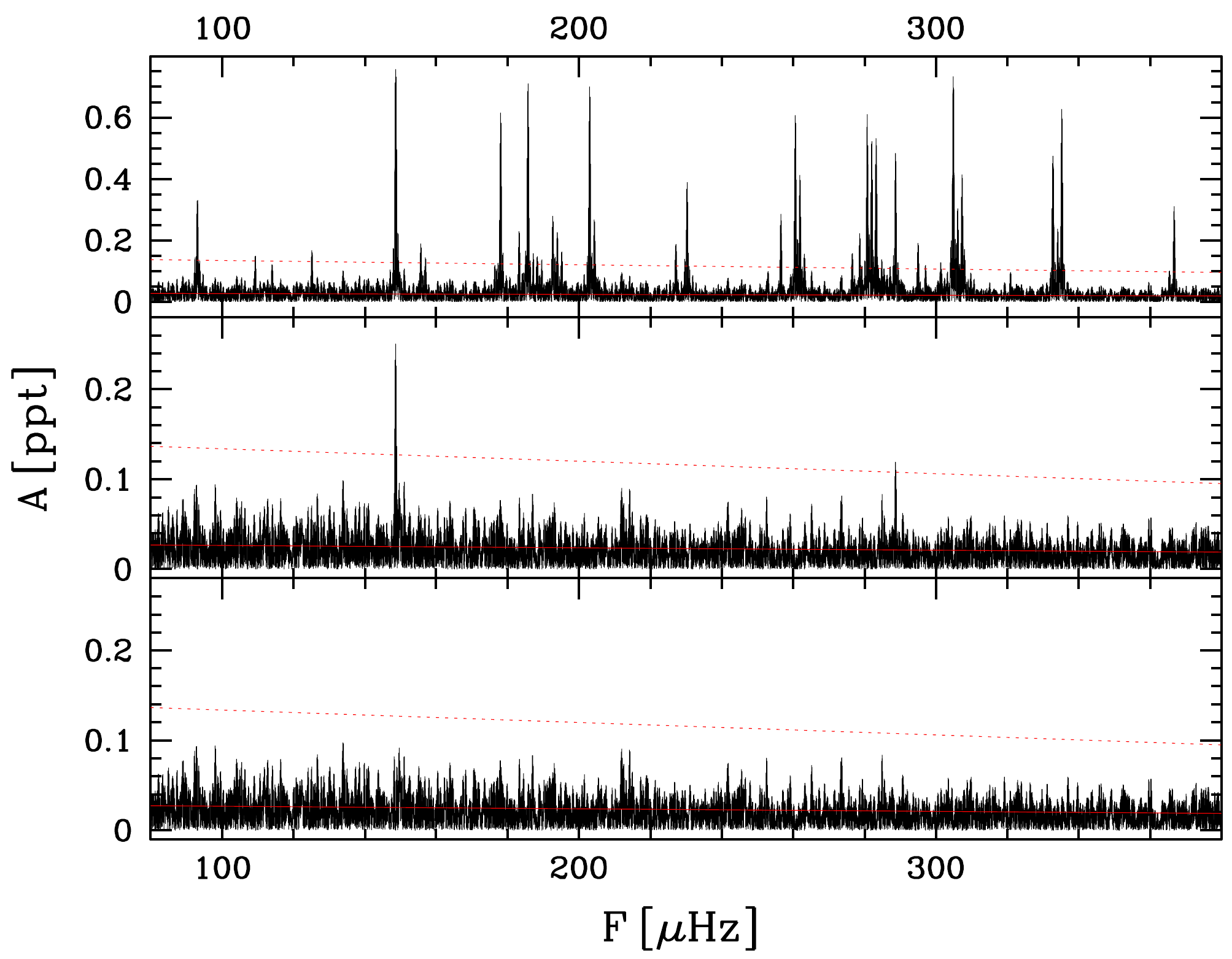}
\vspace{-5mm}
\caption{Prewhitening procedure. Upper panel: spectrum of TIC\,137608661 in 
the main pulsation region. Central panel: residuals after having removed 55 
significant peaks. Bottom panel: residuals after having removed also f6b 
and f33b. The solid red line and the dotted red line are the mean noise level 
and the 5$\sigma$ detection threshold, respectively. See text for more 
details.}
\label{dft_3}
\end{figure}

\begin{figure}
\centering
\includegraphics[width=0.48\textwidth,angle=0]{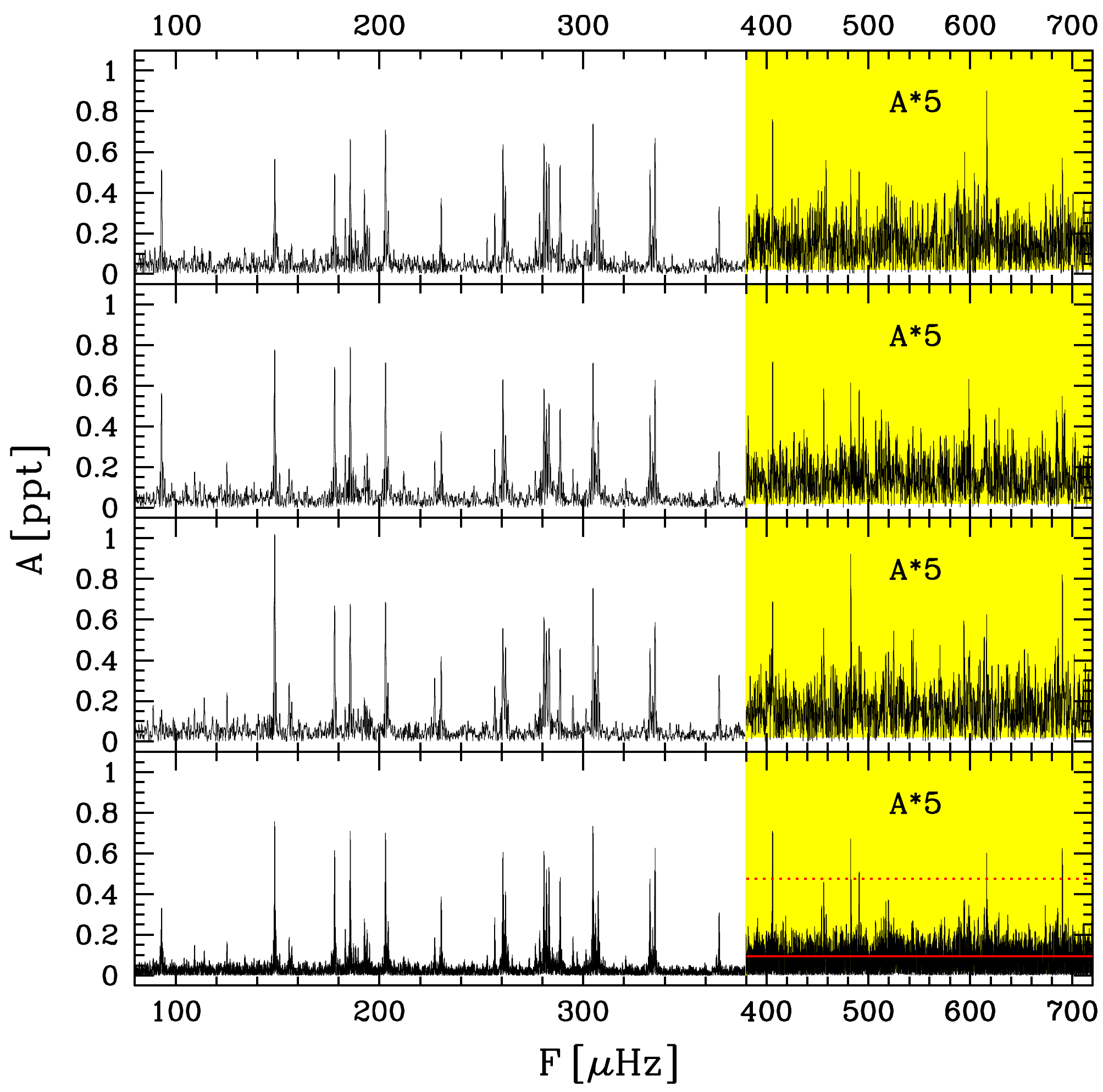}
\vspace{-5mm}
\caption{Amplitude spectrum of each single sector.
From top to bottom: sector 14, 20, 26 and all three sectors together.
Like in Figure~\ref{dft_3}, the solid red line and the dotted red line are the 
mean noise level and the 5$\sigma$ detection threshold, respectively.}
\label{dft_4}
\end{figure}

The Fourier transform (FT) of the $TESS$ data in Figure~\ref{dft} shows a rich 
spectrum.
The low-frequencies are dominated by the orbital modulation at 38.526
$\mu$Hz, with an amplitude of 28.77 ppt, plus its harmonics at 77.052, 115.578 
and 154.105 $\mu$Hz.
As expected, these harmonics have decreasing amplitudes: 4.74, 0.42 and 
0.26 ppt respectively.
A fourth harmonic at 192.645 $\mu$Hz has an amplitude of 0.27 ppt, higher than 
expected, and indeed, as we will see, this frequency is also part of a triplet
of pulsation frequencies, suggesting that a tidal-induced mechanism might be
at work in this case.
Resonance between tides and g-modes pulsations is predicted by 
theory \citep{1975A&A....41..329Z,1977A&A....57..383Z,2018MNRAS.481..715P} 
and may have been seen in a few sdB pulsators 
(e.g. \citealt{2011MNRAS.412..371R,2014A&A...570A.130S}).

At frequencies higher than 90 $\mu$Hz, up to $\sim$370 $\mu$Hz, many peaks
with amplitudes of several hundreds of ppm (part per million) correspond to a
typical sdB spectrum of g-mode pulsations.
A few lower-amplitude peaks ($\sim$100 ppm) are present also at higher 
frequencies, up to 690 $\mu$Hz.

To define a reliable noise threshold we proceeded as follows.
First we computed the signal-to-noise ratio (S/N) corresponding to a False 
Alarm Probability (FAP) of 0.1$\%$ following \citet{Kepler_1993}:
S/N=ln(n$_f$*1000)$^{0.5}$ in which n$_f$=F$_N$/R$_f$ is the number of 
independent frequencies, F$_N$ is the Nyquist frequency, R$_f$=1/$\Delta$T is 
the nominal frequency resolution and $\Delta$T is the total duration of the 
run.
We obtain S/N=4.5.
Then, to test the reliability of this number, we computed the FT of 1000 
simulated light curves obtained by reshuffling in a random way the data, after 
having removed 57 significant frequencies (those in Table~\ref{freqs}).
Since reshuffling destroys any coherent signal, the S/N ratio of the highest 
peak of each FT was used to test the previous S/N expression.
We found that the expression is valid as long as an offset is used and this 
offset is independent from the number of simulations suggesting that it is 
robust.
For this data set the offset on S/N is +0.5 or +0.8 depending whether we use 
the FT mean noise or the FT median noise as the denominator.
These numbers are very similar to those found by \citet{2021arXiv210609718B} 
in their simulations.
In conclusion, using the FT mean noise, we adopted S/N=5.0 as our threshold
for real pulsation frequencies, while the peaks with 4.5$<$S/N$<$5.0 are 
considered as candidates only.
Once the signal-to-noise threshold has been set, we must consider that the mean
noise is not constant everywhere. 
Since it is flat at high frequencies but tends to increase at low frequencies
(due to low-amplitude peaks, below the threshold, unresolved peaks, aliasing 
effects, etc., which are not removed by prewhitening),
after some measurements in different frequency intervals we decided to adopt
a noise model which is linearly decreasing between 90 and 380 $\mu$Hz
(0.027 ppt at 90 $\mu$Hz and 0.019 ppt at 380 $\mu$Hz) and remains constant 
at 0.019 ppt up to 700 $\mu$Hz.

We applied to the light curve a standard prewhitening procedure with nonlinear
least-squares fitting and obtained a list of frequencies that is shown 
in Table~\ref{freqs}.
Those with S/N between 4.5 and 5.0, that are considered as candidates only,
are marked with parentheses in Table~\ref{freqs}.
The prewhitening procedure is illustrated in Figure~\ref{dft_3}.
After prewhitening, only two frequencies were not completely removed from the
amplitude spectrum (central panel of Figure~\ref{dft_3}), leaving residuals 
that could be due either to pairs of very close frequencies below the 
frequency resolution, or to frequency instability (frequencies and/or 
amplitudes that vary over time).

When we compare the amplitude spectra of the various sectors
(Figure~\ref{dft_4}), we see that most peaks are rather stable in amplitude,
despite the fact that from one sector to another there are about 
six months, for a total duration of the observations of about one year.

\begin{table*} 
\centering
\caption[]{Orbital and pulsation frequencies.\\
(Errors in brackets are relative to the last digits, e.g.
38.526250\,(11) means 38.526250$\pm$0.000011).
%{\bf ASB: Roberto, I would definitively work our the error precision. 
%e.g. an error of 294 has no meaning.}
}
\begin{tabular}{lllrrrlr}
\hline
ID & ~~~~F ($\mu$Hz) & ~~~~~~P (s) & A (ppt)~\, & S/N & n$^1$ & $l$ & m \\
\hline
f$_{\rm orb}$   & \hspace{0.6mm} 38.526250\,(11) & 25956.3283\,(72) & 28.774\,(17)\\ 
2*f$_{\rm orb}$ & \hspace{0.6mm} 77.052472\,(65) & 12978.169\,(11)  &  4.738\,(17)\\ 
3*f$_{\rm orb}$ & 115.57770\,(73)  & \hspace{0.6mm} 8652.188\,(55)  &  0.420\,(17)\\ 
4*f$_{\rm orb}$ & 154.1055\,(12)   & \hspace{0.6mm} 6489.062\,(50)  &  0.261\,(17)\\
\hline 
f1      & \hspace{0.6mm} 93.0623\,(10)  & 10745.49\,(12) & 0.297\,(17) &11.0 & & & \\ 
f2      & \hspace{0.6mm} 93.15969\,(99) & 10734.26\,(11) & 0.311\,(17) &11.6 & & & \\ 
f3      & 109.3401\,(21)  & \hspace{0.6mm} 9145.78\,(18)   & 0.144\,(17) & 5.4 &    &	 &     \\ 
(f4     & 114.0211\,(25)  & \hspace{0.6mm} 8770.31\,(19)   & 0.123\,(17) & 4.7)&    &	 &     \\ 
f5      & 125.1936\,(19)  & \hspace{0.6mm} 7987.63\,(12)   & 0.162\,(17) & 6.2 &    &	 &     \\
f6      & 148.63137\,(40) & \hspace{0.6mm} 6728.055\,(18)  & 0.764\,(17) &30.1 & 26 & 1? & +1?\\
f6b$^2$ & 148.6551\,(12)  & \hspace{0.6mm} 6726.982\,(56)  & 0.249\,(17) & 9.8 &    &	 &     \\ % NOT RESOLVED
f7  	& 155.6834\,(17)  & \hspace{0.6mm} 6423.292\,(71)  & 0.179\,(17) & 7.1 & 25 & 1  &  0?\\
f8  	& 156.9764\,(23)  & \hspace{0.6mm} 6370.385\,(92)  & 0.136\,(17) & 5.4 & 25 & 1  & +1?\\
f9  	& 178.03636\,(49) & \hspace{0.6mm} 5616.830\,(16)  & 0.622\,(17) &25.3 & 22 & 1? & +1?\\
f10 	& 183.2003\,(17)  & \hspace{0.6mm} 5458.506\,(50)  & 0.183\,(17) & 7.5 & 21 & 1  & -1 \\
f11 	& 185.69940\,(43) & \hspace{0.6mm} 5385.047\,(13)  & 0.710\,(17) &29.1 & 21 & 1  & +1 \\
f12 	& 188.3160\,(20)  & \hspace{0.6mm} 5310.222\,(57)  & 0.152\,(17) & 6.3 &    &	 &	\\
f13 	& 189.5612\,(22)  & \hspace{0.6mm} 5275.343\,(62)  & 0.137\,(17) & 5.6 &    &	 &	\\
f14 	& 192.6449\,(11)  & \hspace{0.6mm} 5190.898\,(31)  & 0.270\,(17) &11.2 & 20 & 1  & -1 \\
f15 	& 193.9138\,(14)  & \hspace{0.6mm} 5156.930\,(37)  & 0.223\,(17) & 9.2 & 20 & 1  &  0 \\
f16 	& 195.1421\,(19)  & \hspace{0.6mm} 5124.472\,(50)  & 0.162\,(17) & 6.7 & 20 & 1  & +1 \\
f17 	& 203.00214\,(44) & \hspace{0.6mm} 4926.056\,(11)  & 0.693\,(17) &29.0 & 19 & 1  & -1?\\
f18 	& 204.2614\,(12)  & \hspace{0.6mm} 4895.688\,(29)  & 0.257\,(17) &10.8 & 19 & 1  &  0?\\
f19 	& 227.0400\,(17)  & \hspace{0.6mm} 4404.510\,(32)  & 0.185\,(17) & 8.0 &    & 2? &     \\
(f20 	& 227.2388\,(27)  & \hspace{0.6mm} 4400.657\,(53)  & 0.113\,(17) & 4.9)&    &	 &     \\
f21 	& 230.30385\,(78) & \hspace{0.6mm} 4342.090\,(15)  & 0.392\,(17) &16.9 & 17 & 1  & +1?\\
(f22 	& 252.9085\,(30)  & \hspace{0.6mm} 3953.999\,(46)  & 0.104\,(17) & 4.6)&    &	 &     \\
f23 	& 256.6026\,(12)  & \hspace{0.6mm} 3897.077\,(18)  & 0.265\,(17) &11.8 &    & 2? &     \\
f24 	& 260.61413\,(52) & \hspace{0.6mm} 3837.0905\,(77) & 0.591\,(17) &26.5 & 15 & 1  & -1 \\
f25 	& 261.2016\,(21)  & \hspace{0.6mm} 3828.461\,(30)  & 0.150\,(17) & 6.7 &    & 2? &     \\
f26 	& 261.86840\,(70) & \hspace{0.6mm} 3818.712\,(10)  & 0.440\,(17) &19.8 & 15 & 1  &  0 \\
(f27 	& 263.1036\,(29)  & \hspace{0.6mm} 3800.785\,(42)  & 0.105\,(17) & 4.7 & 15 & 1  & +1) \\
f28 	& 276.5385\,(22)  & \hspace{0.6mm} 3616.134\,(28)  & 0.142\,(17) & 6.5 &    & 2  &    \\
f29 	& 278.6794\,(14)  & \hspace{0.6mm} 3588.353\,(18)  & 0.221\,(17) &10.1 &    & 2  &    \\
f30 	& 280.75348\,(51) & \hspace{0.6mm} 3561.8436\,(65) & 0.601\,(17) &27.6 & 14 & 1  & -1 \\
f31 	& 282.00815\,(58) & \hspace{0.6mm} 3545.9968\,(73) & 0.530\,(17) &24.4 & 14 & 1  &  0 \\
f32 	& 283.23899\,(57) & \hspace{0.6mm} 3530.5874\,(71) & 0.540\,(17) &24.9 & 14 & 1  & +1 \\
f33 	& 288.68769\,(65) & \hspace{0.6mm} 3463.9510\,(79) & 0.470\,(17) &21.8 &    & 2? &   \\
f33b$^3$& 288.7149\,(25)  & \hspace{0.6mm} 3463.624\,(30)  & 0.122\,(17) & 5.7 &    &	 &   \\ % NOT RESOLVED
f34  	& 294.9645\,(16)  & \hspace{0.6mm} 3390.238\,(19)  & 0.188\,(17) & 8.8 &    & 2  & -1 or -2\\  
f35  	& 297.0378\,(25)  & \hspace{0.6mm} 3366.575\,(28)  & 0.122\,(17) & 5.7 &    & 2  &  0 or -1\\
f36  	& 301.3289\,(29)  & \hspace{0.6mm} 3318.632\,(31)  & 0.108\,(17) & 5.1 &    & 2  & +2 or +1\\
f37  	& 304.84395\,(40) & \hspace{0.6mm} 3280.3669\,(43) & 0.764\,(17) &36.3 & 13 & 1  & -1 \\
f38  	& 306.1024\,(11)  & \hspace{0.6mm} 3266.880\,(11)  & 0.288\,(17) &13.7 & 13 & 1  &  0 \\
f39  	& 307.33246\,(65) & \hspace{0.6mm} 3253.8053\,(69) & 0.472\,(17) &22.5 & 13 & 1  & +1 \\
(f40  	& 320.8888\,(31)  & \hspace{0.6mm} 3116.344\,(30)  & 0.099\,(17) & 4.8 &    & 2?)&   \\
f41  	& 332.78156\,(70) & \hspace{0.6mm} 3004.9742\,(63) & 0.439\,(17) &21.6 & 12 & 1  & -1 \\
f42  	& 334.0450\,(15)  & \hspace{0.6mm} 2993.608\,(13)  & 0.211\,(17) &10.4 & 12 & 1  &  0 \\
f43  	& 335.27579\,(52) & \hspace{0.6mm} 2982.6192\,(46) & 0.595\,(17) &29.4 & 12 & 1  & +1 \\
(f44  	& 365.4018\,(32)  & \hspace{0.6mm} 2736.714\,(24)  & 0.095\,(17) & 4.9 & 11 & 1  &  0?)\\  
f45  	& 366.69996\,(99) & \hspace{0.6mm} 2727.0251\,(74) & 0.311\,(17) &16.1 & 11 & 1  & +1?\\
f46  	& 405.9003\,(22)  & \hspace{0.6mm} 2463.659\,(13)  & 0.142\,(17) & 7.5 & 10 & 1  & -1?\\
(f47  	& 456.0976\,(34)  & \hspace{0.6mm} 2192.513\,(16)  & 0.091\,(17) & 4.8 &  9 & 1  & -1?)\\
f48  	& 482.5889\,(23)  & \hspace{0.6mm} 2072.157\,(10)  & 0.132\,(17) & 6.9 &    &	&    \\
f49  	& 490.8909\,(30)  & \hspace{0.6mm} 2037.113\,(13)  & 0.102\,(17) & 5.4 &    &	&    \\
f50  	& 616.0114\,(26)  & \hspace{0.6mm} 1623.3466\,(68) & 0.119\,(17) & 6.3 &  7 & 1  & +1?\\
f51  	& 690.3358\,(25)  & \hspace{0.6mm} 1448.5704\,(52) & 0.124\,(17) & 6.5 &    &	&    \\
\hline  
\multicolumn{8}{l}{$^1$ Arbitrary offset. Assuming that the period spacing is 
constant, we consider n=1 for the}\\
\multicolumn{8}{l}{\hspace{2mm} shortest period.}\\
\multicolumn{8}{l}{$^2$ Residual (unresolved) peak after removing f6. 
See text for more details.}\\
\multicolumn{8}{l}{$^3$ Residual (unresolved) peak after removing f33. 
See text for more details.}\\
\end{tabular}	    
\label{freqs}	    
\end{table*}	    

\begin{figure*}
\centering
\includegraphics[width=1.008\textwidth,angle=0]{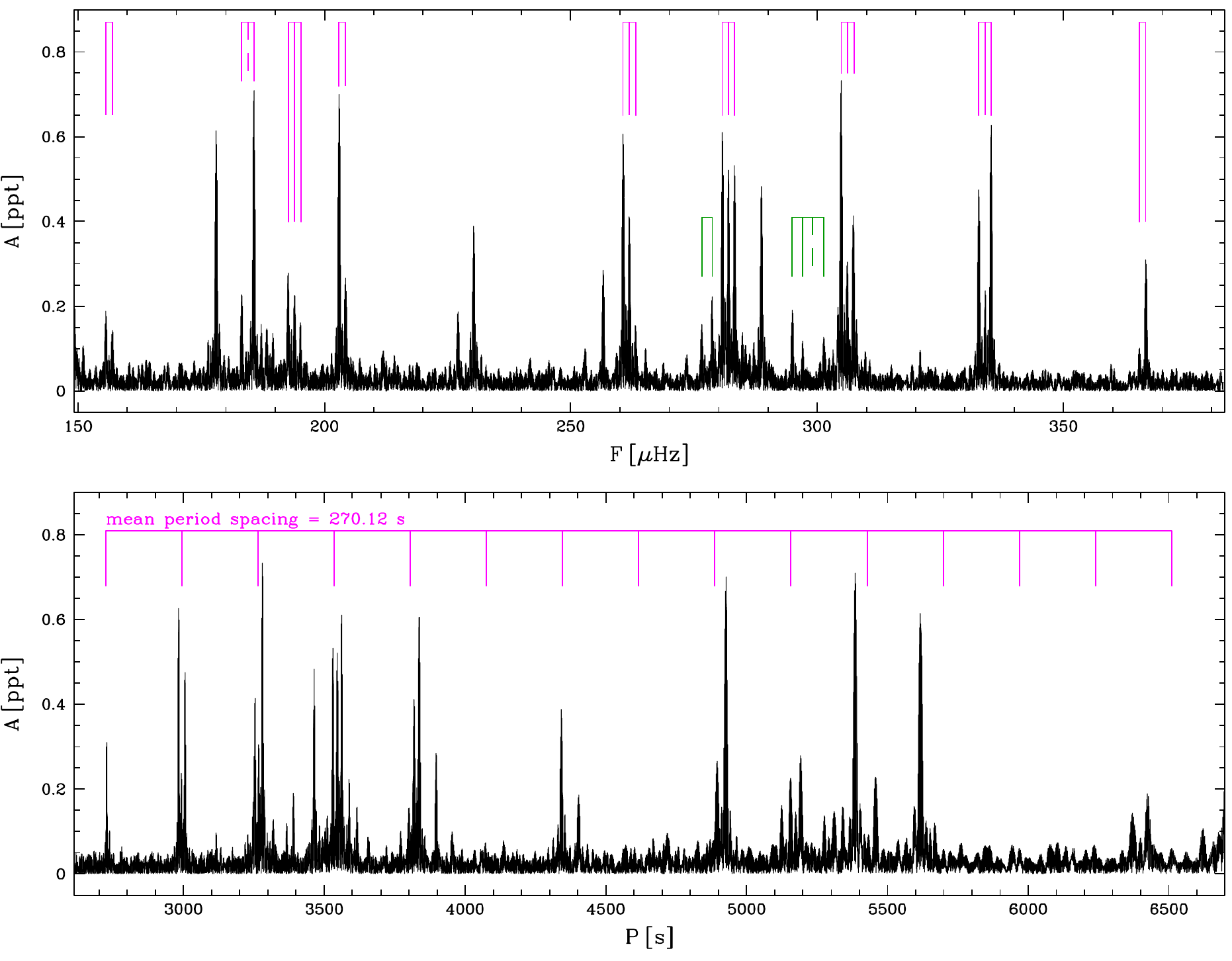}
\vspace{-5mm}
\caption{Frequency splitting (top) vs period spacing (bottom).
Magenta refers to $l$=1 modes while green refers to $l$=2.
Both panels represent exactly the same region of the spectrum.}
\label{dft_freq_splitting}
\end{figure*}

\subsection{Period spacing, frequency splitting, inclination of the rotation 
axis}

\begin{figure}
\centering
\includegraphics[width=0.48\textwidth,angle=0]{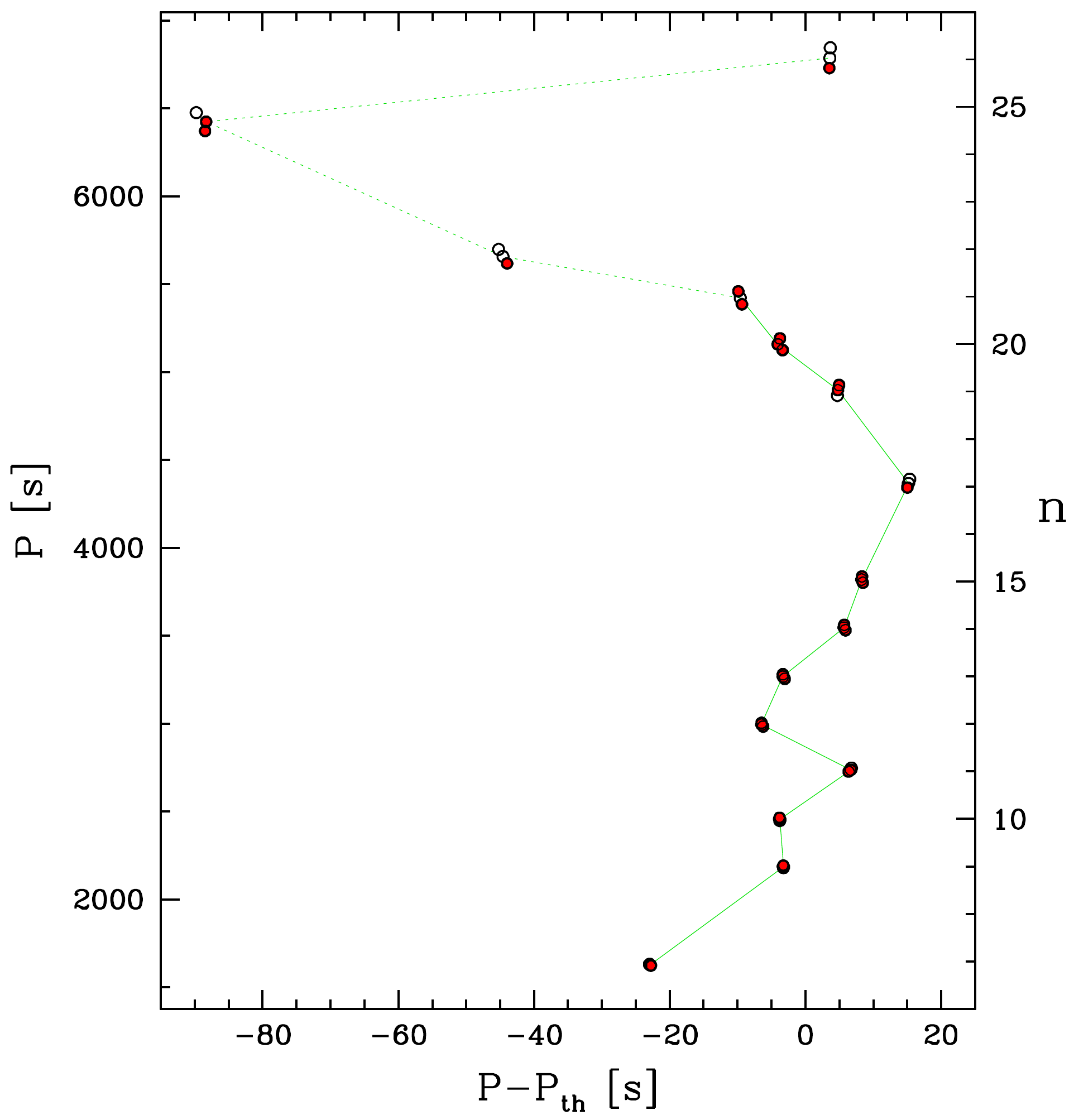}
\vspace{-5mm}
\caption{\'Echelle diagramme of the dipole modes showing the residuals
between observed periods and theoretical periods computed assuming a constant
period spacing of 270.116 s. The open {\bf circles} are the missing components
of the rotational triplets.}
\label{echelle}
\end{figure}

\begin{figure}[ht]
\centering
\includegraphics[width=0.48\textwidth,angle=0]{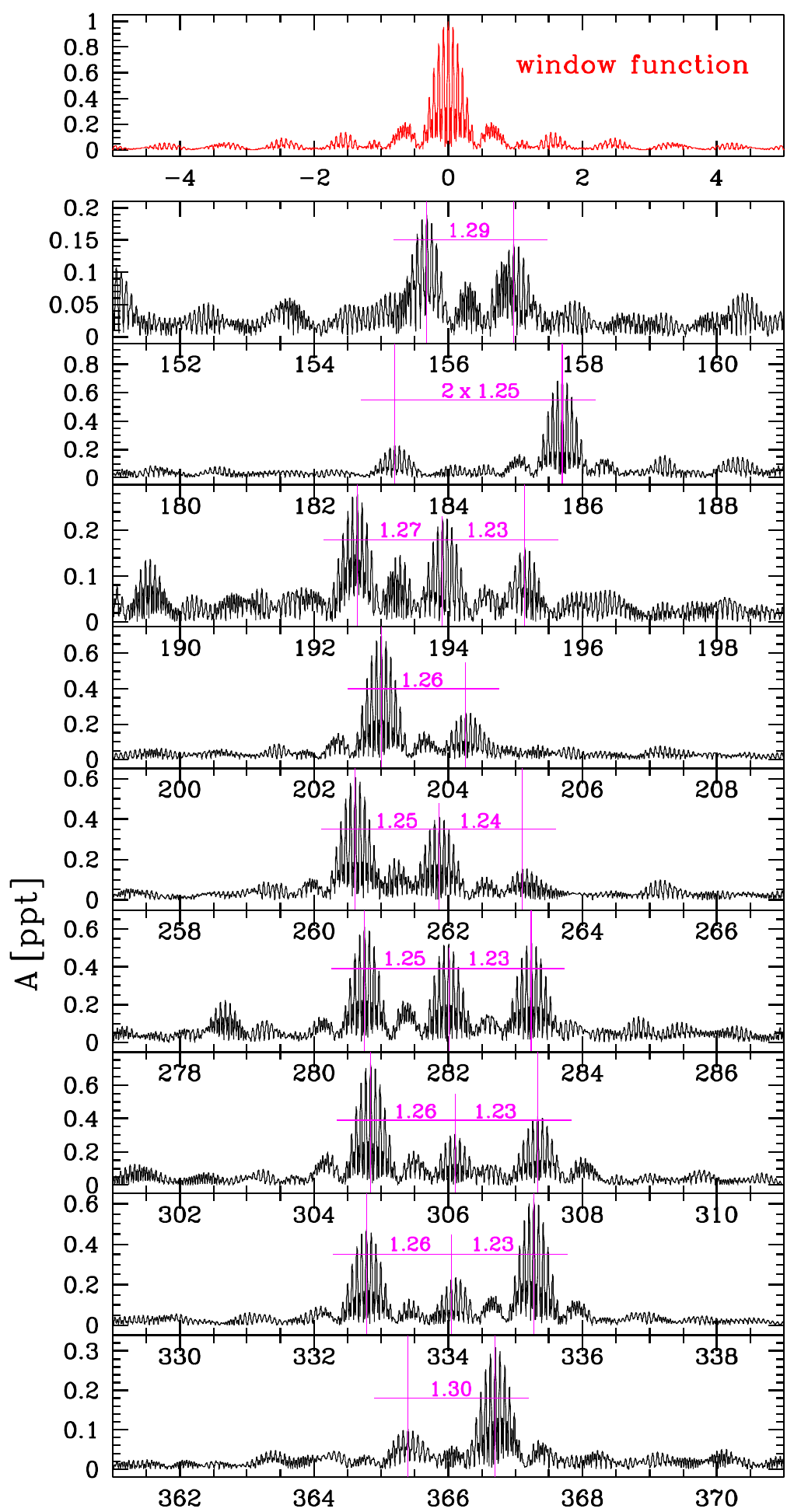}
\includegraphics[width=0.48\textwidth,angle=0]{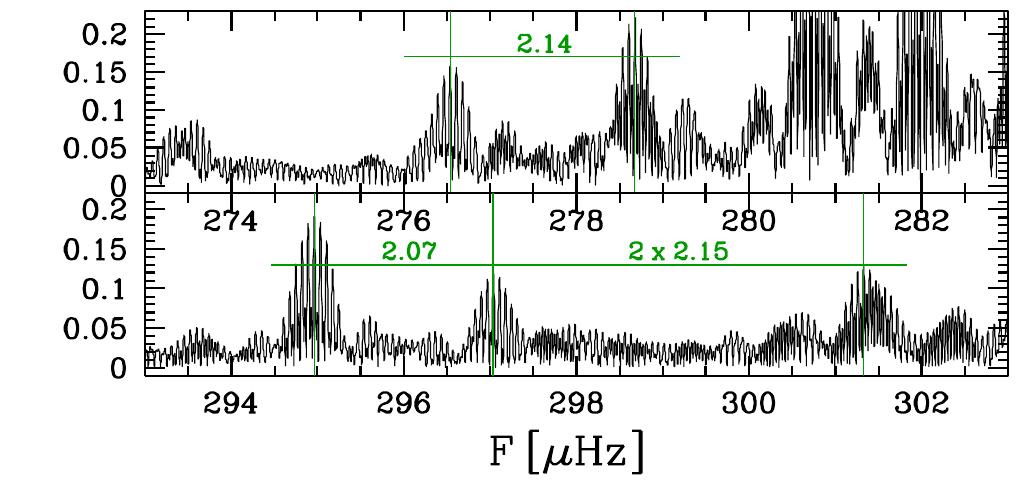}
\vspace{-5mm}
\caption{Frequency splitting in detail: the nine complete or
incomplete $l$=1 triplets and the two incomplete $l$=2 quintuplets.}
\label{dft_10}
\end{figure}

\begin{figure}
\centering
\includegraphics[width=0.48\textwidth,angle=0]{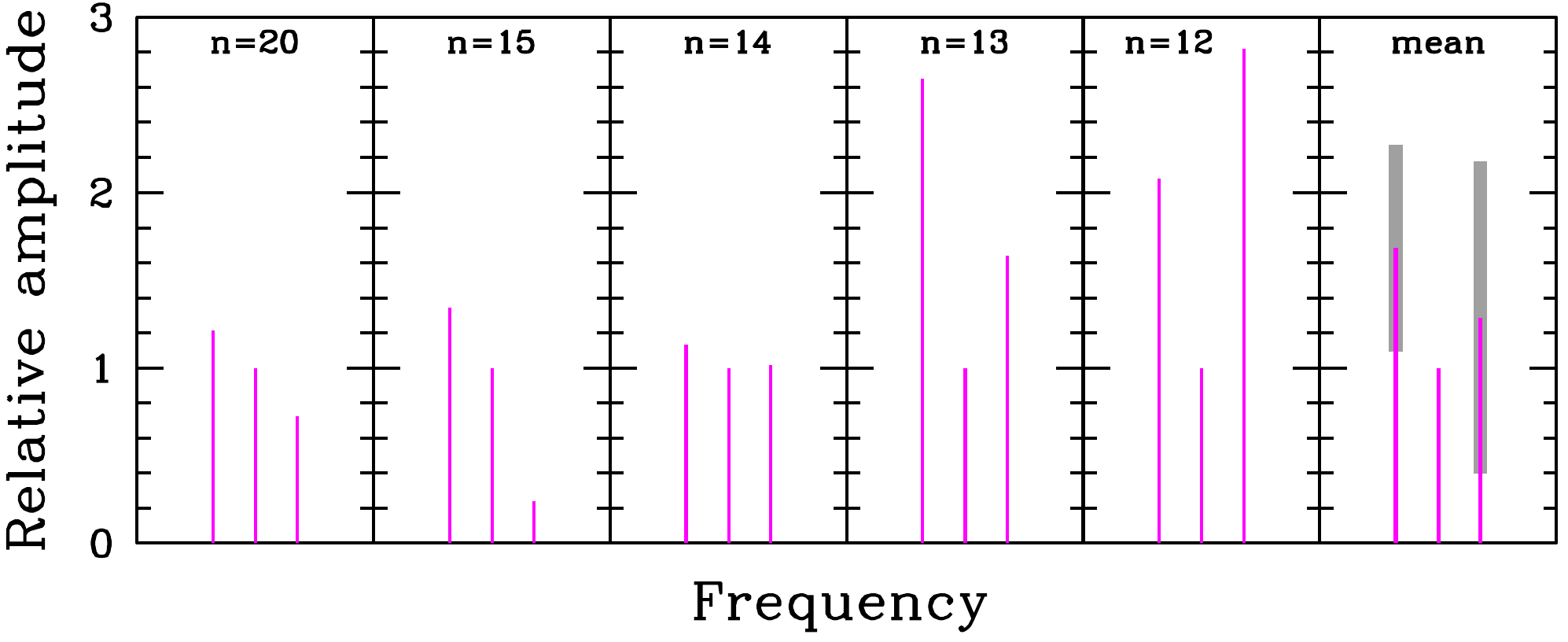}
\vspace{-5mm}
\caption{Relative amplitudes of the components of the five $l$=1 triplets. 
The amplitude of the central component of each triplet is normalized to 1 
before computing the mean amplitudes, which are shown in the rightmost panel.
The error bars (in grey) are the standard deviations around the mean 
amplitudes.}
\label{inclination}
\end{figure}

Looking at the amplitude spectrum of TIC\,137608661, 
%(in particular the upper panel of Figure~\ref{dft_3}, which has a larger scale
%in frequency), 
we immediately note four consecutive triplets of equally spaced frequencies
between $\sim$260 and $\sim$340 $\mu$Hz with a frequency spacing of about 
1.3 $\mu$Hz (Figure~\ref{dft_freq_splitting} upper panel).
Another triplet located near $\sim$190 $\mu$Hz and a few doublets 
between $\sim$150 and $\sim$370 $\mu$Hz show a very similar frequency spacing.
When we plot the same region of the spectrum in the period domain
(lower panel of Figure~\ref{dft_freq_splitting}), 
we see that the central peaks of the four well defined triplets
between 260 and 340 $\mu$Hz are equally spaced in period.
The period difference between the central peaks at 3818.7, 3546.0, 3266.9
and 2993.6\,s is 272.7, 279.1 and 273.3\,s respectively, confirming that the
triplets are consecutive $l$=1 modes.
The mean period spacing of these four triplets is 275.03\,s.
Since the central peak of the triplet near $\sim$190 $\mu$Hz,
which has a period of 5156.9\,s, is also well compatible with the $l$=1 
sequence of modes, we include also this period in the computation of the mean 
period spacing. From a linear least-squares fit to the five $m$=0 periods we 
obtain $\Delta$P$=270.12\pm1.19$\,s.
Using this value, several other doublets fall close to the expected periods 
(assuming a perfectly constant spacing) and therefore we adopt 270.12\,s 
in our analysis.
Once the period spacing is fixed, we were able to identify further modes, 
including a few $l$=1 single peaks at short and long periods.
The geometry of the identified modes (number of radial nodes n, 
spherical degree $l$, and azimuthal quantum number m) is reported in 
Table~\ref{freqs}.
In Figure~\ref{echelle} the ``\'echelle diagramme'' of the $l$=1 sequence shows
the residuals between observed and theoretical periods.
We note the typical meandering shape between $\sim$2000 and $\sim$5000-6000\,s
that is also seen in other sdB pulsators such as 
KIC\,10553698A \citep{Ostensen_2014b},
EPIC\,211779126 \citep{Baran_2017}, 
KIC\,10001893 \citep{2017MNRAS.472..700U},
KIC\,11558725 \citep{2018MNRAS.474.4709K},
PHL\,457 \citep{2019MNRAS.489.1556B}.
And we see that the periods at n=22 (f9) and n=25 (f7 and f8) are 
significantly shorter than the expected values from a constant period spacing,
indicating a possible mode trapping.
Although an incorrect identification as dipole modes can not be totally 
excluded, the frequency difference between f7 and f8, equal to 1.29 $\mu$Hz, 
suggests that f7 and f8 are indeed two components of the same $l$=1 triplet.

Figure~\ref{dft_10} shows all the complete or incomplete $l$=1 triplets of 
frequencies split by the rotation of the star.
Considering all of them, we obtain a mean frequency splitting of 1.254 $\mu$Hz
corresponding to a rotation period of about 4.6 days in the deep layers 
of the star.\\
The rotation period is obtained from the expression 
P$_{\rm rot}$=(1--C$_{nl}$)/$\delta\sigma_{nl}$ (where $\delta\sigma_{nl}$
is the frequency spacing), which is valid for a slowly rotating star with
$\Omega_{\rm rot}<<\sigma_{\rm puls}$.
For high-order g-modes, in the asymptotic limit, we can use the following 
approximation:
C$_{n,l}$ $\simeq$\,1/[($l$\,($l$+1)] and we obtain C$_{(l=1)}$$\simeq$1/2 and 
C$_{(l=2)}$$\simeq$1/6 \citep{Ledoux_1951, Unno_1989, Aerts_2010}.
Although a sequence of $l$=2 equally spaced periods is not seen in our data,
from the $l$=1 splitting of 1.254 $\mu$Hz we can compute 
the expected frequency splitting for the $l$=2 modes:
$\delta\sigma_{(l=2)}$\,$\simeq$\,$\frac{5}{3}\,\delta\sigma_{(l=1)}$$\simeq$\,2.090 $\mu$Hz.
And indeed we see two multiplets of frequencies (a doublet and a 
triplet) which have frequency separations very close to this number.
These incomplete quintuplets are shown in the lower panels of 
Figure~\ref{dft_10} and are reported in Table~\ref{freqs}.
The $l$=2 spherical degree has been tentatively attributed also to a few other
frequencies.

There is another valuable piece of information that can be derived from
the very clean amplitude spectrum of TIC\,137608661.
Having a certain number of clearly identified $l$=1 triplets, we can analyze
the amplitude of each m-component to derive an estimate of the inclination
of the rotation axis respect to the line of sight.
More precisely, the geometric visibility of each m-component, and therefore 
its amplitude, depends on the angle between the pulsation axis and the line 
of sight.
However, misalignments between pulsation and rotation axis, which would
split each $l$=1 mode in nine components \citep{Pesnell_1985},
have never been seen in sdB stars.
Thus we can safely assume that pulsation and rotation axes are aligned.
If we also assume that, on average, each m-component of a multiplet
receive the same amount of energy and develop approximately the same 
intrinsic amplitude level, then the mean amplitude ratio between m=$\pm$1
and m=0 components of a certain number of triplets should directly reflect the 
inclination of the rotation axis. 
With five well-defined triplets, this measurement should already have some 
level of accuracy.
In Figure~\ref{inclination} we show the relative amplitudes of the components
of the five triplets, the mean amplitudes of the m=--1,0,+1 components, 
and their errors.
A comparison between these numbers and those computed by 
\citet[][Supplementary Information, Figure\,A.5; 
see also \citealt{Pesnell_1985}]{2011Natur.480..496C},
allows us to exclude low inclinations and suggests an inclination of 
(65$^{+10}_{-20}$)$\degr$ for the rotation axis of the sdB star.

With such inclination, and adopting a radius of 0.209~\rsun\ from the SED
analysis (section 5), the projected equatorial rotation velocity near the 
surface of the star $V_{\rm e}\sin{i}$\,=\,7.5 km/s obtained in section 4 
would correspond to an orbital period of 1.3 days, shorter than the 4.6 
days obtained in the deep layers, suggesting a differential rotation,
although a rigid rotation can not be totally ruled out as discussed 
in section 4.

\subsection{Asteroseismic analysis}

\begin{figure}
\centering
\includegraphics[width=0.5\textwidth,angle=0]{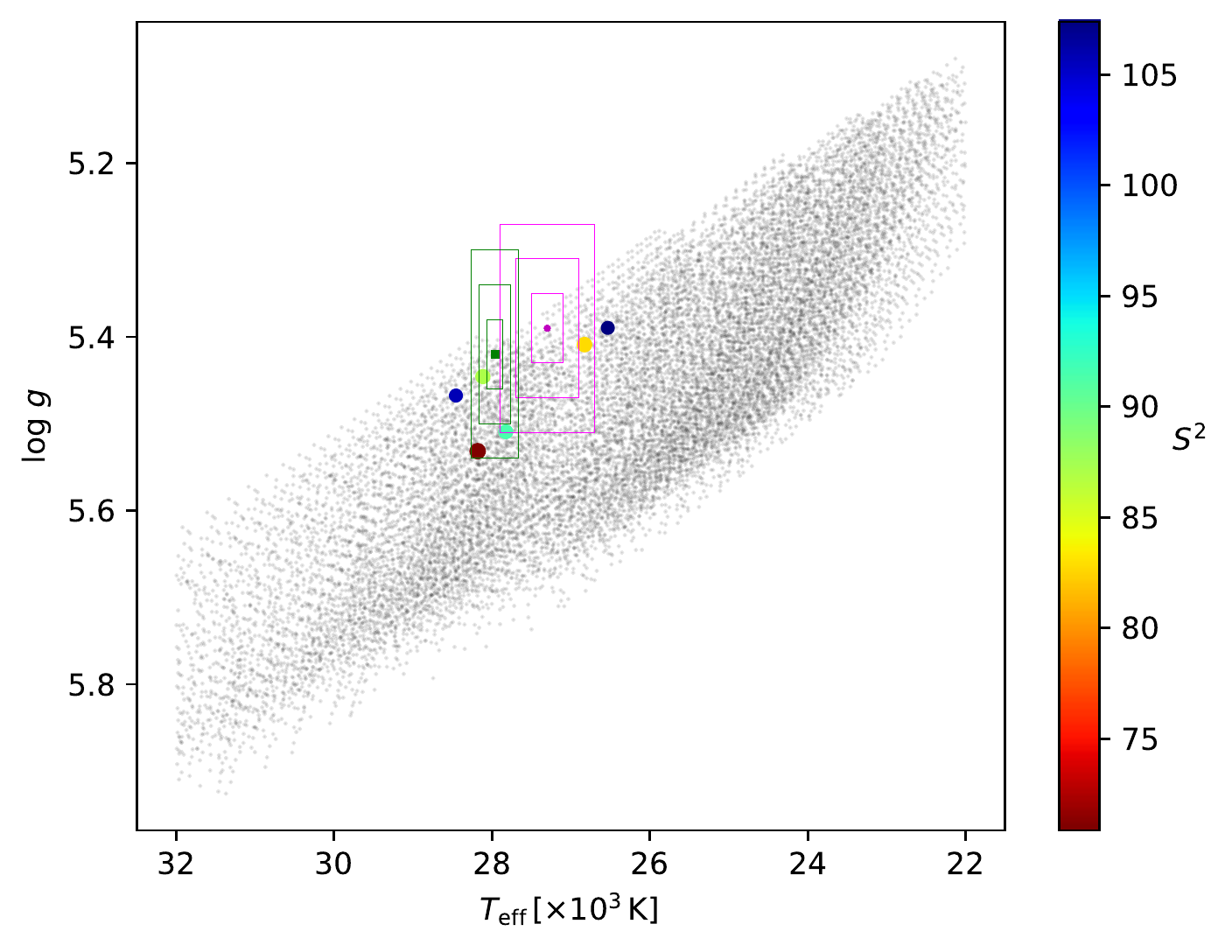}
\vspace{-5mm}
\caption{Asteroseismic determination of \teff\ and \logg.
Color coded $S^2$ (as defined in equation~\ref{eq:1}) outlines the best 
models. Minimum $S^2$ is obtained for \teff=28180\,K and \logg=5.53
(dark-red point). Magenta and dark-green small points/rectangles represent
spectroscopic \teff\ and \logg\ with 1, 2 and 3\,$\sigma$ error boxes, from 
LTE and non-LTE models respectively.}
\label{astero_analysis}
\end{figure}

In order to obtain more insight about evolutionary status and interior of 
TIC\,137608661, we calculated evolutionary models using the \texttt{MESA} code 
\citep[Modules for Experiments in Stellar 
Astrophysics;][]{Paxton_2011,Paxton_2013,Paxton_2015,Paxton_2018,Paxton_2019}, 
version 11701. The models were calculated for progenitors with initial masses, 
$M_\mathrm{i}$, in the range of $1.0\,-\,1.8\,M_\odot$, with a step of 
$0.01\,M_\odot$, and metallicities, $Z$, in the range of $0.005\,-\,0.035$, 
with a step of $0.005$. The initial helium abundance was determined by the 
linear enrichment law, $\Delta\,Y/\Delta\,Z=1.5$. The protosolar helium 
abundance, $Y_{\odot,\,\mathrm{protosolar}}=0.2703$, and the mixture of metals 
were adopted from \citet{2009ARA&A..47..481A}. The progenitors were evolved to 
the tip of the red giant branch where, before the helium ignition, most of the 
hydrogen has been removed leaving only a residual hydrogen envelope on top of 
the helium core. The considered envelope masses, M$_\mathrm{env}$, are in the 
range of $0.0001\,-\,0.0030\,M_\odot$, with a step of $0.0001\,M_\odot$. The 
models were then relaxed to an equilibrium state and evolved until the 
depletion of helium in the core. In all calculations we used the novel 
convective premixing scheme in order to ensure proper growth of the convective 
core during the course of evolution \citep{Paxton_2019}. 
The thorough description of the models is provided in 
\citet{2021MNRAS.503.4646O}. The adiabatic 
pulsation calculations were performed using the GYRE code, version 5.2 
\citep{Townsend_2013,Townsend_2018}. The pulsation models were calculated for 
evolutionary models with central helium abundance, $Y_\mathrm{c}$, in the range
of $0.9\,-\,0.1$, with a step of $0.05$. The models with $Y_\mathrm{c}<0.1$ 
were not considered due to the occurrence of the breathing pulses, which are 
unavoidable side effects of the convective premixing scheme 
\citep{2021MNRAS.503.4646O}.

The grid of evolutionary models was used to find the models that represent 
TIC\,137608661. No spectroscopic constraints were used in the process and we 
have chosen to fit only the five pulsation periods corresponding to the m=0
dipole modes identified via multiplet structures (f15, f26, f31, f38 and f42
in Table~\ref{freqs}).
We used a goodness-of-fit function, which calculates the difference between  
observed and theoretical periods
\begin{equation}
S^2=\frac{1}{N_{\rm o}}\,\sum^{N_{\rm o}}_{i=1} \left(P_{\rm o}^i-P_{\rm c}^i\right)^2
\label{eq:1}
\end{equation}
where $P_\mathrm{o}$ is an observed period, $P_\mathrm{c}$ is a calculated 
period, and $N_\mathrm{o}$ is the number of periods used (5 in this case). 
The minimum of the $S^2$ function indicates the best fit. 

%Here, we are only considering solutions that have $S^2$ up to about $150\%$ 
%of the obtained minimum.
%
%While we did not obtained a single unique solution for the star, the 
%considered

Considering only the six best solutions with $S^2$ up to about $150\%$ of the 
minimum $S^2$, these best models clearly indicate that the progenitor of 
TIC\,137608661 is a star with a metallicity close to solar, $Z=0.01 - 0.015$, 
with a mass $M_\mathrm{i}=1.1-1.2\,M_\odot$. 
The estimated age of the star is $5.4-7.3$ Gyr. 
The envelope mass of the star is constrained in the narrow range
$M_\mathrm{env}=0.0006-0.0009\,M_\odot$.
%In all the six best models the sdB mass is close to 0.47 \msun\ (0.473-0.477),
%and the radius is between 0.19 and 0.23 \rsun.
As shown in Figure~\ref{astero_analysis}, the best solutions are located close 
to the spectroscopic determinations of $T_\mathrm{eff}$ and $\log\,g$. 
Such spectroscopic values suggest a rather evolved EHB model, which 
is supported by the low central helium abundance of the best models, 
$Y_\mathrm{c}=0.3-0.1$.

%%%%%%%%%%%%%%%%%%%%%%%%%%%%%%%%%%%%%%%%%%%%%%%%%%%%%%%%%%%%%%%%%%%%%%%%%%%%%%

\section{TIC\,137608661 in context: synchronized vs non-synchronized sdBs 
in short-period binaries}

\begin{table*} 
\centering
\caption[]{SdB rotation (from asteroseismology) vs orbital periods.}
\begin{tabular}{lllllr}
\hline
Name$^*$ & P$_{\rm orb}$ (d) & P$_{\rm rot}$ (d) & P$_{\rm rot}$ (d) & Comments & Ref. \\
     &                   & g-modes           & p-modes           &          &\\
\hline
\multicolumn{5}{l}{$\bullet$ sdB+dM binaries}\\
\hline
NY\,Vir (PG\,1336-018)            & 0.10083 &         & 0.10083& rigid rot. down to 0.55 R$_{\star}$ & (1)\\
%KIC\,9472174 (2M\,1938+4603)     &0.1257653& 0.1258 ?& NO Prot found !! & \citealt{2010MNRAS.408L..51O};\\
%                                 &         &         &                  & \citealt{2015A&A...577A.146B}\\
TYC1\,4544-2658-1 (TIC\,137608661)& 0.300   & 4.6     &        & & (2)\\
PHL\,457 (EPIC\,246023959)        & 0.31289 & 4.6     & 2.5\,? & & (3)\\
V1405\,Ori (EPIC\,246683636)      & 0.39802 & 4.2\,?  & 0.555  & & (4)\\
B4 in NGC\,6791 (KIC\,2438324)    & 0.3985  & 9.20    &        & & (5,6,7)\\
KIC\,11179657		          & 0.394   & 7.4     &        & & (8,6)\\
KIC\,2991403		          & 0.443   &10.3     &        & & (8,6)\\
EQ\,Psc (EPIC\,246387816)         & 0.80083 & 9.4     &        & & (3)\\
\hline
\multicolumn{5}{l}{$\bullet$ sdB+WD binaries}\\
\hline
HD\,265435 (TIC\,68495594)     &  0.0688185&      & 0.069&                & (9)\\
%KL UMa (Feige 48)             &  0.38     &	  & 0.38 & rigid rotation & (9)\\
KL UMa (Feige 48)              &  0.34361  &	  & 0.38 & rigid rotation & (10,11)\\
PG\,1142-037 (EPIC\,201206621) &  0.5411   &>45.  &      &	          & (12)\\
KIC\,7664467                   &  1.5590   & 35.1 &      &	          & (13)\\
EPIC\,211696659                &  3.16     &>28.  &      &	          & (14)\\
KIC\,10553698	               &  3.39     & 41.  &      &	          & (15)\\
KIC\,11558725	               & 10.05     & $\sim$45.&  $\sim$40.& $\sim$44 d rigid rotation & (16,17)\\
FBS\,1903+432 (KIC\,7668647)   & 14.17     & 46-48 & 49-52\,? & close to rigid rotation & (18)\\
\hline
\multicolumn{5}{l}{$\bullet$ sdB+FGK wide binaries}\\
\hline
PG\,0048+091 (EPIC\,220614972) & ?   & 13.9\,? &  4.4   & sdB+F   & (19)\\
EPIC\,211823779                & ?   &         & 11.5\,?& sdB+F1V & (14)\\
PG\,1315-123 (EPIC\,212508753) & ?   & 15.8    & 16.2   & sdB+F   & (19)\\
EGGR\,266 (EPIC\,211938328)    & 635 &         & 21.5\,?& sdB+F6V & (14)\\
\hline
\multicolumn{5}{l}{$\bullet$ single sdBs}\\
\hline
%BAL\,0090100001               & & 6.3 ?? non so da dove venga questo valore  & & & \citealt{2009MNRAS.392.1092B}\\
EPIC\,211779126                & &         & $\sim$16 & core rotation likely slower & (20)\\
UY\,Sex (EPIC\,248411044)      & &         & 24.6     &         		    & (4)\\
KIC\,10139564	               & &$\sim$26 & $\sim$26 & rigid rotation              & (21,22)\\
KY UMa (PG\,1219+534)	       & &         & 34.9     & rigid rot. down to 0.6 R$_{\star}$ & (23)\\
KPD\,1943+4058 (KIC\,5807616)  & &  39.2   &          & 			    & (24)\\
KIC\,2697388	               & &$\sim$42 & $\sim$53 & close to rigid rotation	    & (25,26)\\
KIC\,3527751                   & &  42.6   & 15.3\,?  &                             & (27,28)\\
EPIC\,203948264  	       & &$\ge$45.9&          & 			    & (29)\\
TIC\,33834484	               & &  64     & & 					    & (30)\\
EPIC\,212707862  	       & &$\sim$80 & & 					    & (31)\\
KIC\,10670103		       & &  88     & & 					    & (32)\\
KIC\,1718290	               & &$\sim$100& & BHB star\,!			    & (33)\\
KIC\,10001893	               & & 289     & & 					    & (34)\\
\hline
\multicolumn{6}{l}{$^*$ Kepler/K2/TESS id are used as 1st or 2nd name when the results are based on Kepler/K2/TESS data.}\\
\multicolumn{6}{l}{(1) \citealt{Charpinet_2008}; (2) this paper; (3) \citealt{2019MNRAS.489.1556B}; 
(4) \citealt{2020MNRAS.492.5202R}; (5) \citealt{2011ApJ...740L..47P};}\\
\multicolumn{6}{l}{(6) \citealt{2012AcA....62..343B}; (7) \citealt{2022MNRAS.509..763S};
(8) \citealt{2012MNRAS.422.1343P}; (9) \citealt{Pelisoli_2021};}\\
\multicolumn{6}{l}{(10) orbital period from TESS data preliminary analysis; (11) \citealt{VanGrootel+2008}; 
(12) \citealt{2016MNRAS.458.1417R};}\\ 
\multicolumn{6}{l}{(13) \citealt{Baran_2016}; (14) \citealt{2018MNRAS.474.5186R}; (15) \citealt{Ostensen_2014b}; 
(16) \citealt{Telting_2012};}\\
\multicolumn{6}{l}{(17) \citealt{2018MNRAS.474.4709K}; (18) \citealt{Telting_2014}; (19) \citealt{2019MNRAS.483.2282R}; 
(20) \citealt{Baran_2017};}\\ 
\multicolumn{6}{l}{(21) \citealt{2012MNRAS.424.2686B}; (22) \citealt{Zong_2016}; (23) \citealt{2018OAst...27...44V}; 
(24) \citealt{2011Natur.480..496C} (SI);}\\ 
\multicolumn{6}{l}{(25) \citealt{2012AcA....62..179B}; (26) \citealt{2017MNRAS.465.1057K}; 
(27) \citealt{2015ApJ...805...94F}; (28) \citealt{2018ApJ...853...98Z}; (29) \citealt{2017MNRAS.467..461K};}\\ 
\multicolumn{6}{l}{(30) Uzundag et al. in prep.; (31) \citealt{2016AcA....66..455B}; 
(32) \citealt{2014MNRAS.440.3809R}; (33) \citealt{2012ApJ...753L..17O};}\\ 
\multicolumn{6}{l}{(34) \citealt{Charpinet_2018}.}\\
\end{tabular}	    
\label{rot_vs_orb}	    
\end{table*}	    

With an orbital period of 7.21 hours, a well defined rotation period 
of 4.6 days in the deep layers, and a lower limit to the
rotation period of about 1.3 days near the surface, 
the sdB star in TIC\,137608661 is relatively far from synchronization.
However, among the handful of sdB+dM binaries for which the sdB rotation was 
measured through asteroseismology, it is the non-synchronized system with
both the shortest orbital period and the shortest sdB rotation period.
Only PHL\,457, with basically the same sdB core rotation period and a slightly 
longer orbital period of 7.51 hours \citep{2019MNRAS.489.1556B}, 
ranks at the same level in the synchronization process.
Table~\ref{rot_vs_orb} shows the list of sdB/sdO stars for which the
rotation period was measured from g- or p-mode frequency splitting.

A few stars are not reported in Table~\ref{rot_vs_orb}:\\
{\bf $\bullet$ KIC\,2991276} since it is not clear whether this star is single 
or in a binary system. The short rotation period (6.3~d) measured from p-modes
\citep{Ostensen_2014a} suggests the presence of a companion and the low
amplitude of the m=$-$1,+1 modes belonging to the triplets near 7560 and
8200 $\mu$Hz suggest a low inclination $i$\lsim15$\degr$.
A low inclination, together with KIC\,2991276's faintness (17th-mag),
means that it is not easy to verify the presence of a companion 
for this star.\\
{\bf $\bullet$} Also {\bf Balloon\,090100001} shows a similar rotation period 
between 6.4 and 7.4 days from p-modes, but in this star the frequency 
splitting changes in 1 year \citep{2009MNRAS.392.1092B}. 
The reasons for this peculiar behaviour are unclear and the presence of
a magnetic field with the strength of $\sim$1\,kG was excluded by
\citet{2013ARep...57..751S}.\\
{\bf $\bullet$} 
No rotational splitting was detected in {\bf HD\,4539/EPIC\,220641886} and 
{\bf KIC\,8302197} \citep{Silvotti_2019,2015A&A...573A..52B}, suggesting 
long rotation periods (too long to be measured respect to the observing runs)
and/or low inclinations.\\
{\bf $\bullet$}
{\bf B3} and {\bf B5} in the open cluster NGC\,6791 show a mean rotation 
period of 64.2 and 71 days respectively 
(the latter somewhat uncertain), for both stars RV variations suggest 
the presence of a companion, but the orbital periods are unknown
\citep{2022MNRAS.509..763S}.
Interestingly, in B3 the l=1 g-mode frequency splitting decreases at
increasing frequencies, suggesting that differential rotation could
operate even on a small radius scale.\\
{\bf $\bullet$}
No frequency multiplets were found in the amplitude spectrum of
{\bf 2M\,1938+4603}, an sdB+dM binary with an orbital period 
of 3.02 hours \citep[and references therein]{2015A&A...577A.146B}.\\
{\bf $\bullet$}
A clear detection of rotational splitting is missing in
{\bf KPD\,1930+2752}, an sdB+WD binary with an orbital period of 
only 2.3 hours that shows ellipsoidal variations
\citep{2000ApJ...530..441B,2011MNRAS.412..371R}.\\
With such short orbital periods, both the sdB stars in
2M\,1938+4603 and KPD\,1930+2752 should be synchronized
or very close to synchronization.\\

\begin{figure}
\centering
\includegraphics[width=0.48\textwidth,angle=0]{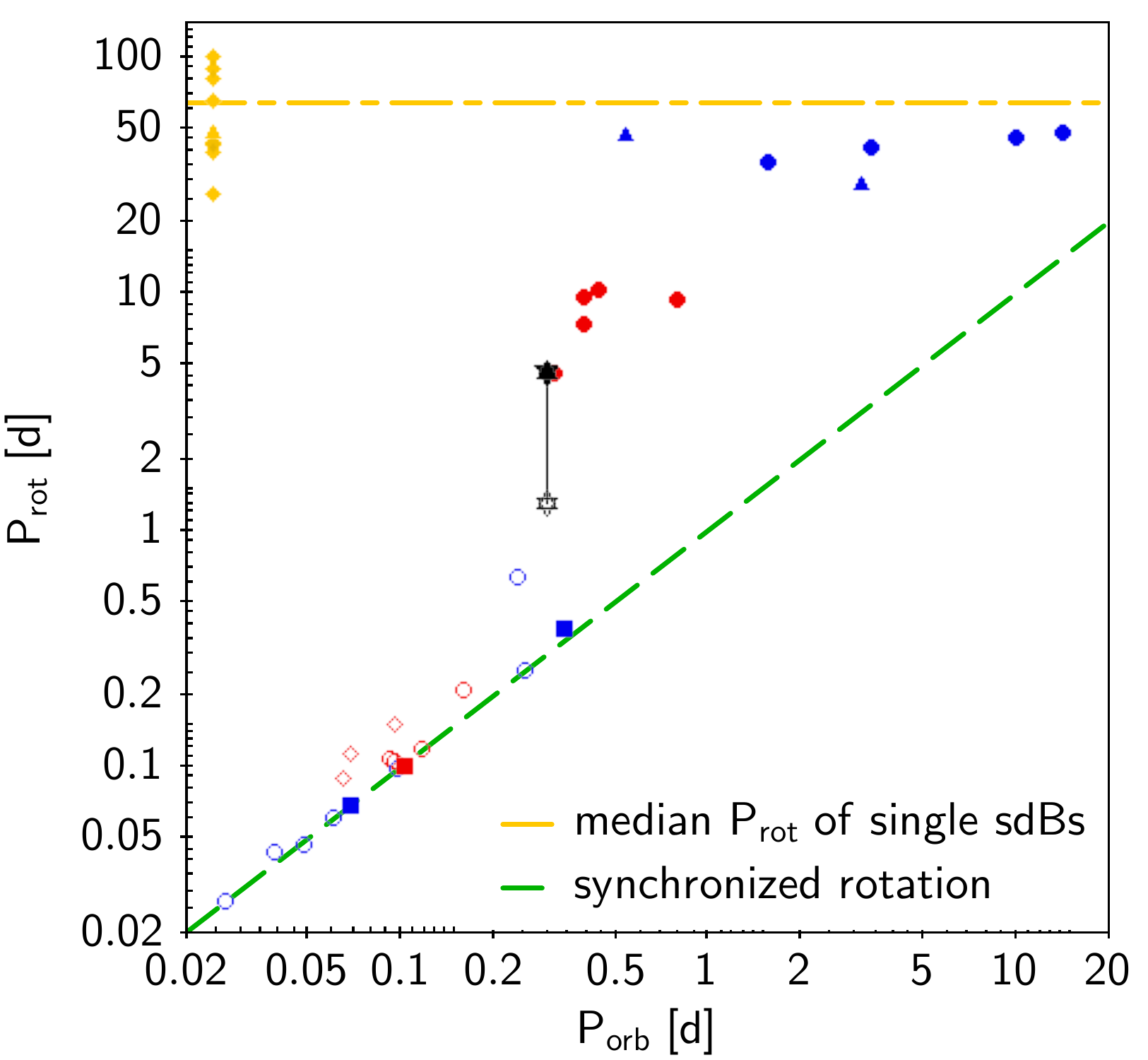}
\vspace{-3mm}
\caption{SdB rotation period vs orbital period.
Red symbols: sdB+dM systems with sdB rotation period obtained from
g-mode or p-mode frequency splitting (filled circles and filled square
respectively) or from spectral line broadening (empty circles or empty 
diamonds, the latter indicating a brown dwarf companion).
Black symbols: TIC\,033834484, deep-layers rotation period from g-modes 
frequency splitting (filled asterisk), and lower limit to 
the surface rotation period from spectral line broadening 
(empty asterisk).
Blue symbols: sdB+WD systems with sdB rotation period obtained from
g-mode frequency splitting (filled circles or filled triangles that
indicate a lower limit) or p-mode frequency splitting (filled squares)
or from spectral line broadening (empty circles).
Yellow symbols: single sdB stars with rotation period obtained 
from g-mode frequency splitting (triangle indicating a lower limit).
The seven red empty symbols correspond to the following HW Virginis 
(eclipsing sdB+dM/BD) systems: from left to right
V\,2008-1753 \citep{Schaffenroth_2015,2021MNRAS.501.3847S},
SDSS\,J162256.66+473051.1 \citep{Schaffenroth_2014},
PTF1\,J011339.09+225739.1 \citep{2018OAst...27...80W},
HS\,0705+6700 \citep{2001A&A...379..893D},
SDSS\,J082053.53+000843.4 \citep{2021MNRAS.501.3847S},
HW\,Vir \citep{2010A&A...519A..25G,2018MNRAS.481.2721B,2021A&A...648A..85E},
EPIC\,216747137 \citep{Silvotti_2021}.
The seven blue empty circles correspond to the following non-eclipsing
(or partially eclipsing) sdB+WD binaries:
from left to right
ZTF\,J2130+4420 \citep{Kupfer_2020a},
ZTF\,J2055+4651 \citep{Kupfer_2020b},
CD-30$\degr$11223 \citep{2012ApJ...759L..25V},
PTF1\,J0823+0819 \citep{2017ApJ...835..131K},
EVR-CB-001 \citep{2019ApJ...883...51R},
PG\,2345+318 \citep{2010A&A...519A..25G,2004Ap&SS.291..267G},
EVR-CB-004 \citep{2020ApJ...902...92R}.
}
\label{synchro}
\end{figure}

Focusing on short-period binaries (both sdB+dM and sdB+WD systems), 
Figure~\ref{synchro} includes also some systems for which the sdB rotation
period is inferred from the rotational velocity measured through spectral 
line broadening:
P$_{\rm rot}$=2$\pi$R\,sin\,$i$/(v\,sin\,$i$).
This technique is more efficient at very short rotation periods 
and high equatorial velocities,
when the slow-rotation approximation normally used for the rotational splitting
may no longer be valid.
This might be the reason why we do not see frequency splitting in
2M\,1938+4603 and KPD\,1930+2752.

We see from Figure~\ref{synchro} that synchronization occurs for orbital 
periods shorter than $\sim$0.3 days (first group of stars), while at orbital 
periods longer than $\sim$1 day (second group) the rotation periods are close 
to the typical values of single stars, of a few tens of days.
Between these two groups, a third group is formed by the stars that are 
approaching synchronization.
It is important to note that different methods were used to measure 
P$_{\rm rot}$ in these three groups and these methods sample different 
regions of the star: deep layers with the g-mode frequency splitting
used in group 2 and 3, external layers with the spectral line broadening 
or p-mode frequency splitting used in group 1.
Indeed, a few systems in Table~\ref{rot_vs_orb}, for which the sdB rotation 
was measured at different depths, suggest that differential rotation
might be quite common in these stars.
And this can partially explain why the jump between group 2 and 3 and 
group 1 that we see in Figure~\ref{synchro} is so steep.

Figure~\ref{synchro} suggests a few further comments:
in group 1 the three systems that have a brown dwarf (BD) companion 
(represented with red empty diamonds) are not fully synchronized,
differently from the other sdB+dM systems with similar orbital periods.
Although we do not know their evolutionary phase, this effect may be
related to their longer synchronization time which is inversely proportional 
to the companion mass.
At longer orbital periods, the two sdB/sdO+WD systems {\bf PG\,2345+318} and 
{\bf EVR-CB-004} (blue circles in Figure~\ref{synchro}), with almost identical 
orbital periods of 0.24 and 0.25~d respectively, have significantly different 
rotation periods.
But EVR-CB-004 hosts a peculiar object with a radius of 0.63 \rsun\
that can be either an inflated sdO star or, more likely, a post-blue 
horizontal branch star \citep{2020ApJ...890..126R}.
With such a radius, it is not surprising that the star was most affected
by the tidal effects from its WD companion. 
PG\,2345+318, on the other hand, is a key object in a transition region 
between non-synchronized and synchronized systems. We know that it must have 
an inclination close to 90$\degr$ because \citet{2004Ap&SS.291..267G} saw a 
primary (and may be a secondary) eclipse in the light curve. 
Thus, from v$_{\rm rot}$\,sin$i$=12.9~km/s and \logg=5.70 
\citep[and references therein]{2010A&A...519A..25G} and assuming 
$i$=90$\degr$ and M=0.47~\msun, we obtain P$_{\rm rot}$$\simeq$0.63~d.
Another interesting system, {\bf PG\,1232-136}, with an orbital period of 
0.363~d and a very low rotational velocity v$_{\rm rot}$\,sin$i$<5 km/s 
\citep[and references therein]{2010A&A...519A..25G},
is not represented in Figure~\ref{synchro} since the unknown inclination
leaves two different possibilities open: a) the system is not synchronized and 
the companion is likely a white dwarf; b) the system is synchronized, the
inclination must be very low ($i$<14$\degr$), and the companion 
is a black hole candidate.
From a preliminary analysis of the $TESS$ data, the light curve of PG\,1232-136
shows a weak (730 ppm) orbital modulation at exactly the orbital period.

%%%%%%%%%%%%%%%%%%%%%%%%%%%%%%%%%%%%%%%%%%%%%%%%%%%%%%%%%%%%%%%%%%%%%%%%%%%%%%

\section{Summary}

TIC\,033834484 is a new sdB+dM binary with an orbital period of 7.21 hours.
The $TESS$ light curve shows the typical orbital modulation produced by the 
heating of the secondary star and shows also a rich spectrum of g-mode 
pulsations from the primary.

The atmospheric parameters of the sdB star are well compatible with the 
g-mode instability strip. 
From 13 low-resolution spectra collected with ALFOSC@NOT we obtain
\teff=27300$\pm$200\,K, \logg=5.39$\pm$0.04,
log${\rm (N_{He}/N_{H})}$=--2.95$\pm$0.05
from LTE or \teff=27960$\pm$100\,K, \logg=5.42$\pm$0.04,
log${\rm (N_{He}/N_{H})}$=--2.89$\pm$0.05 from non-LTE models, 
respectively\footnote{But LTE \teff\ is obtained from only 3 spectra near 
orbital phase 0, while non-LTE \teff\ is obtained from all the 13 ALFOSC 
spectra, see section 3.}.
The general metal abundance pattern observed in sdB stars is characterized 
by sub-solar light metal abundances while Fe is typically solar 
\citep{2013A&A...549A.110G}.
The chemical abundances from non-LTE models of TIC\,033834484 agree 
with this general pattern, however the star is relatively poor in Si, 
while the Fe abundance is over twice the solar value. 
%\citep[see e.g.][and references therein]{Heber_2016}.

The amplitude spectrum is particularly simple to interpret as we see five
well defined $l$=1 triplets of frequencies in which all the three 
m=--1,0,+1 components are clearly visible. 
A few more incomplete triplets are also present, that allow us
to obtain a $l$=1 mean period spacing $\Delta$P\,=\,270.12\,$\pm$\,1.19~s.
The mean $l$=1 frequency splitting of 1.254 $\mu$Hz
corresponds to a robust rotation period of about 4.6 days
in the deep layers of the star.
From the mean amplitude of the m=--1,+1 modes of the five complete triplets
we can also constrain the inclination of the rotation axis
and we obtain $i$=(65$^{+10}_{-20}$)$\degr$.

The spectroscopic measurements of \teff\ and \logg\ are in good agreement 
with the best values that we obtain from an asteroseismic analysis using 
the \texttt{MESA} code, although the asteroseismic analysis suggests 
a slightly higher surface gravity.
Adiabatic pulsation computations applied to the best evolutionary 
models and compared with the observed pulsation periods, suggest that the 
progenitor of TIC\,033834484's primary was a star with an initial mass 
of 1.1--1.2 \msun, with a solar metallicity.
They suggest also that the sdB star has an envelope mass of 0.0006--0.0009 
\msun\ and is rather evolved, with a central helium abundance of 10--30\%.
Furthermore, they indicate that the total age of the system is 5.4--7.3 Gyr.

Since the SED of TIC\,033834484 does not show any contribution from the 
companion, we can infer that its effective temperature must be lower than 
about 4000~K.
Moreover, from the LTE/non-LTE surface gravity (\logg=5.39 or 5.42 
respectively), and the stellar radius (0.209 \rsun\ from the SED) we obtain 
an sdB mass between 0.39 and 0.42 \msun, that can reach 0.47 \msun\ when 
we consider the best surface gravity and radius resulting from the 
asteroseismic analysis (\logg=5.53 and R=0.196 \rsun).
Then, from P$_{\rm orb}$=0.300 d and K=41.9 km/s, we obtain a companion 
mass M$_{c}$\,sin$i$=83--94 \mjup, not far from the hydrogen burning limit.
Which means a mass of $\sim$93--105 \mjup\ if we assume that 
the sdB rotation axis is inclined $\sim$65$\degr$ to the line of sight 
(as obtained in section 6.2) and is perpendicular to the orbital plane.
Following the low-mass models by \citet{2015A&A...577A..42B}, a mass
of 100 \mjup\ corresponds to a M dwarf with R=0.120 \rsun, \teff=2750~K and
log(L/L$_\odot$)=--3.12 when we consider an age between 5 and 8 Gyr.

The measurement of the rotation period of TIC\,033834484's primary 
in the deep layers of the star is important because only in 
a handful of sdB pulsators the rotational splitting is so well defined, 
leading to a robust determination of the rotation period.
It is particularly interesting because TIC\,033834484 falls in a critical 
and poorly populated region of the P$_{\rm orb}$-P$_{\rm rot}$ plane 
(Figure~\ref{synchro}), in which the sdB star is gaining angular momentum 
without having already reached the full synchronization with the orbital 
period.
For these reasons we tried to measure also the rotation rate in the outer 
layers of the star to verify if the star rotates as a rigid body or not.
From 33 high-resolution spectra collected with HERMES@Mercator 
we can rule out a surface rotation synchronized with the orbital motion.
Although from single line profiles we can not determine an accurate rotation
velocity, however, using many sharp metal lines together, we are able to
obtain a projected equatorial rotation velocity $V_{\rm e}\sin{i}=7.5\pm1.5$ 
km/s, which would correspond to a rotation period of 1.3 days 
assuming an inclination of 65$\degr$.
Even if this result suggests a differential rotation 
for the sdB star in TIC\,033834484, as found in few other sdB stars
\citep[see e.g.][]{2021MNRAS.507.4178R}, 
other phenomena, different from rotation, that are discussed 
in section 4, may contribute to the spectral line broadening observed. 
Therefore 7.5 km/s should be considered as an upper limit to the projected 
rotation velocity and we can not completely rule out a rigid rotation.

To measure the sdB rotation in sdB+dM/sdB+WD short-period binaries is of
considerable importance for guiding theoretical studies on tides
and tidal synchronization time-scales.
As pointed out by \citet{2018MNRAS.481..715P}, sdB synchronization time-scales 
seem to be longer than the sdB lifetime and, in particular, the synchronization
of NY\,Vir remains not explained by current models, even when we consider
a larger convective core \citep{2019MNRAS.485.2889P}.
Potential explanations given by these authors for the synchronization
of NY\,Vir are a partial synchronization 
of at least the outer layers of the star already during the common envelope 
phase, or higher convective mixing velocities respect to those obtained 
with the mixing length theory.

%%%%%%%%%%%%%%%%%%%%%%%%%%%%%%%%%%%%%%%%%%%%%%%%%%%%%%%%%%%%%%%%%%%%%%%%%%%%%%

\section*{Acknowledgements}
This paper is based on photometric data collected by the $TESS$ space 
telescope, low-resolution spectroscopic data collected with the 2.6\,m Nordic 
Optical Telescope (NOT), and high-resolution spectroscopic data collected with 
the 1.2\,m Mercator Telescope.
Both NOT and Mercator are operated on the island of La Palma at the Spanish
Observatorio del Roque de los Muchachos of the Instituto de
Astrof\'isica de Canarias.
NOT is owned in collaboration by the University of Turku and Aarhus University,
and operated jointly by Aarhus University, the University of Turku and the 
University of Oslo, representing Denmark, Finland and Norway, the University 
of Iceland and Stockholm University.
Mercator is operated by the Flemish Community.  
The high-resolution data are obtained with the HERMES spectrograph, which is 
supported by the Research Foundation - Flanders (FWO), Belgium, the Research 
Council of KU Leuven, Belgium, the Fonds National de la Recherche Scientifique 
(F.R.S.-FNRS), Belgium, the Royal Observatory of Belgium, the Observatoire de 
Gen\`eve, Switzerland, and the Th\"uringer Landessternwarte Tautenburg, 
Germany.
Funding for the $TESS$ mission is provided by the NASA Explorer Program.  
Funding for the $TESS$ Asteroseismic Science Operations Centre is provided by 
the Danish National Research Foundation (Grant agreement n. DNRF106), ESA 
PRODEX (PEA 4000119301) and Stellar Astrophysics Centre (SAC) at Aarhus 
University. 
We thank the $TESS$ team and the TASC/TASOC team for their support to the 
present work and in particular we thank St\'ephane Charpinet for organising and
coordinating the TASC Working Group 8 on Evolved Compact Stars.
In this article we also made use of data obtained with the Far Ultraviolet 
Spectroscopic Explorer, through the MAST data archive at the Space Telescope 
Science Institute, which is operated by the Association of Universities for 
Research in Astronomy, Inc., under NASA contract NAS 5-26555.
RS acknowledges financial support from the INAF project on
``Stellar evolution and asteroseismology in the context of the PLATO space 
mission'' (PI S. Cassisi).
PN acknowledges support from the Grant Agency of the Czech Republic 
(GA\v{C}R 18-20083S). 
This research has used the services of \mbox{\url{www.Astroserver.org}} under 
reference UX87S2.
Asteroseismic analysis calculations have been carried out using resources 
provided by Wroclaw Centre for Networking and Supercomputing 
(\url{http://wcss.pl}), grant No. 265.
Financial support from the National Science Centre under projects 
No.\,UMO-2017/26/E/ST9/00703 and UMO-2017/25/B/ST9/02218 is acknowledged.
We thank Uli Heber, J.J. Hermes, Marcelo M. Miller Bertolami and an anonymous 
referee for useful comments.

%%%%%%%%%%%%%%%%%%%%%%%%%%%%%%%%%%%%%%%%%%%%%%%%%%

\section*{Data availability}

The $TESS$ data underlying this article is publicly available, the 
spectroscopic data may be requested to the authors.



\bibliographystyle{mnras.bst}
\bibliography{biblio} % if your bibtex file is called biblio.bib

%% Alternatively you could enter them by hand, like this:
%% This method is tedious and prone to error if you have lots of references

%%\begin{thebibliography}{99}

%%\bibitem[\protect\citeauthoryear
%%{Author}{2012}]{Author2012}
%%Author A.~N., 2013, Journal of Improbable Astronomy, 1, 1
%%
%%\bibitem[\protect\citeauthoryear{Others}{2013}]{Others2013}
%%Others S., 2012, Journal of Interesting Stuff, 17, 198

%\end{thebibliography}

%%%%%%%%%%%%%%%%%%%%%%%%%%%%%%%%%%%%%%%%%%%%%%%%%%

%%%%%%%%%%%%%%%%% APPENDICES %%%%%%%%%%%%%%%%%%%%%

\newpage
\newpage

\appendix

\section{Non-LTE 2D errors on effective temperature and surface gravity}

%\onecolumn

\begin{figure}
\includegraphics[width=\linewidth]{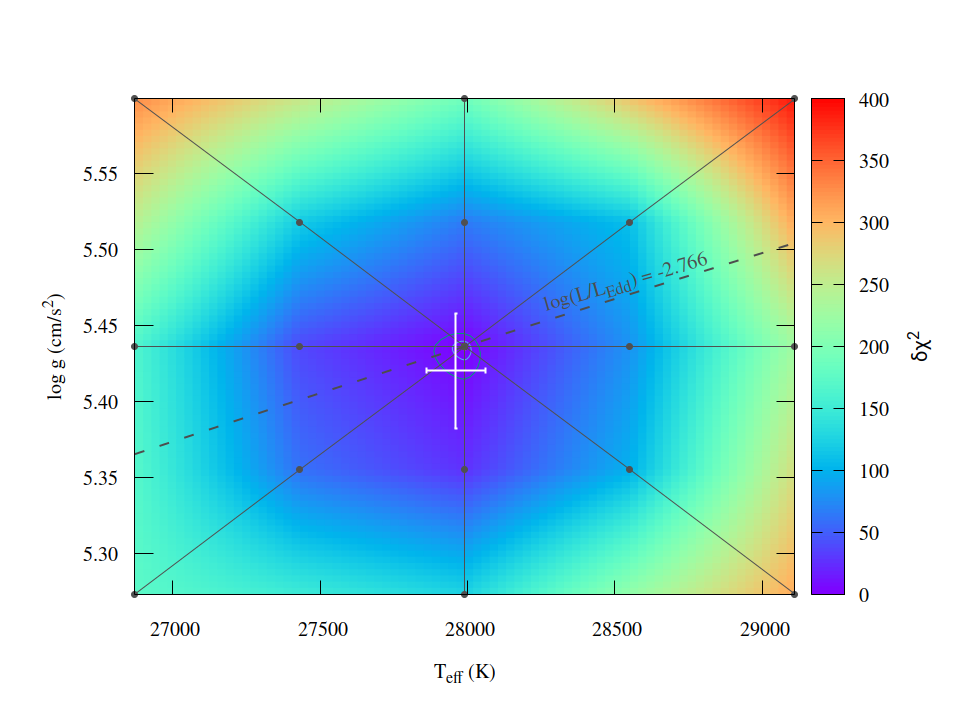}
\vspace{-7mm}
\caption{Non-LTE \teff--\logg\ correlation
and two dimensional error determination for TIC\,137608661. 
The color bar shows the chi-square variations with the parameters. 
The contours are for 60, 90 and 99\% confidence intervals and the white error 
bars represent the adopted final errors.}
\label{errors}
\end{figure}

%\twocolumn
%\newpage

\vspace{16mm}

\section{Fit of the co-added HERMES spectrum}

%\vspace{16mm}

\begin{figure*}
\includegraphics[width=\linewidth]{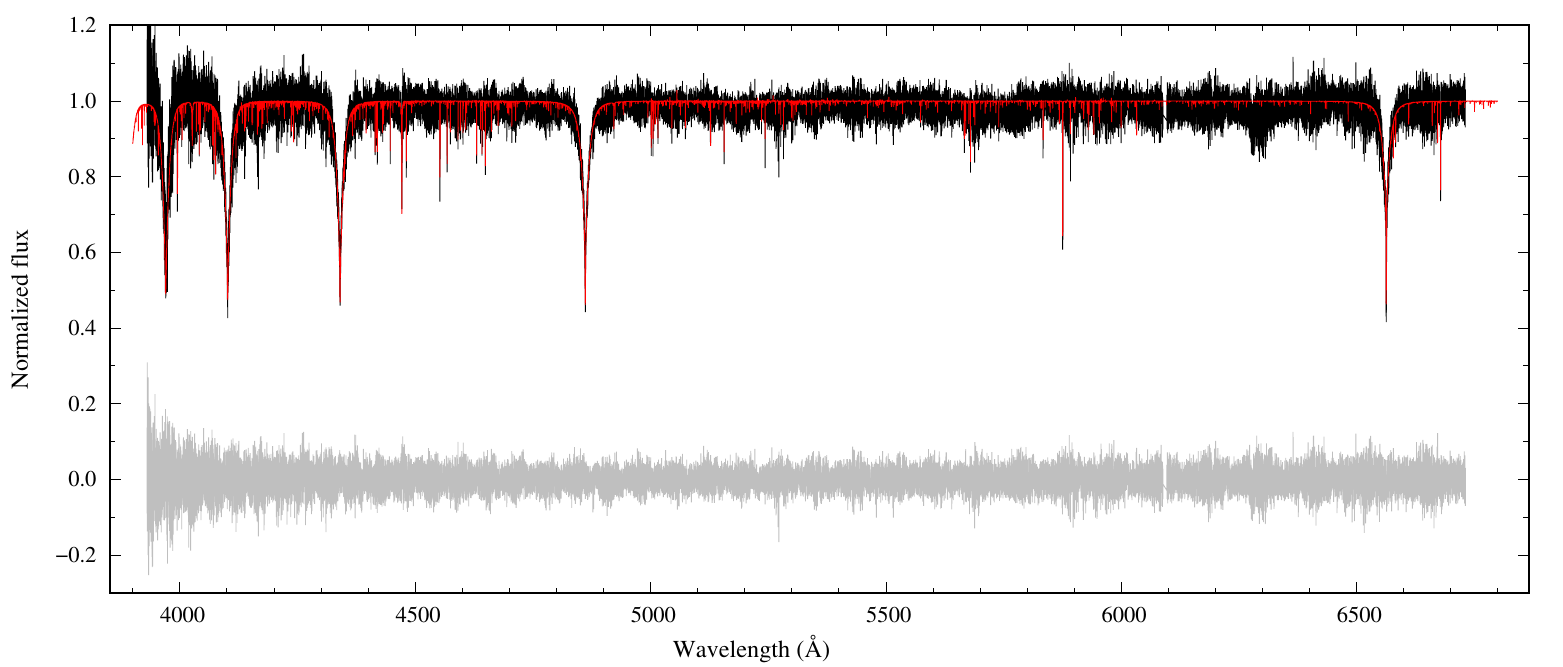}
\vspace{-3mm}
\caption{Fit (and residuals) of the entire HERMES spectrum using {\sc XTgrid}.
The wavelength ranges used to determine the surface rotation velocity are
specified in Table~\ref{WLranges}.}
\label{HR_fit}
\end{figure*}

\begin{table*}
\setstretch{1.1}
\centering
\caption{ 
Wavelength ranges used to determine the surface rotation velocity.}
\label{WLranges}
\begin{tabular}{ll}
\hline 
~~~\,$\lambda$ (\AA) & main lines\\
\hline
3990\,--\,4000 & N\,II~3995.00\\
4020\,--\,4050 & HeI~4026.21; NII~4035.07,\,4041.30,\,4043.53; FeIII~4039.16\\
4060\,--\,4080 & OII~4069.62,\,4069.88,\,4072.15,\,4075.86\\ 
4110\,--\,4130 & OII~4119.22; FeIII~4121.34,\,4122.03,\,4122.78\\ 
4230\,--\,4250 & NII~4236.92,\,4241.79\\ 
4410\,--\,4423 & OII~4414.9,\,4416.97; FeIII~4419.59\\
4442\,--\,4450 & NII~4447.03\\
4465\,--\,4486 & HeI~4471.50; MgII~4481.12,\,4481.32\\
4546\,--\,4651 & SiIII~4552.62,\,4567.83,\,4574.76; OII~4590.96,\,4596.18,\,4638.85,\,4641.81,\,4649.14; NII~4601.48,\,4607.15,\,4630.53,\,4643.09\\
4996\,--\,5052 & NII~5001.13,\,5001.47,\,5005.15,\,5007.33,\,5010.62,\,5045.10; HeI~5015.68\\
5660\,--\,5690 & NII~5666.63,\,5676.02,\,5679.56,\,5686.21\\
5869\,--\,5881 & HeI~5875.61\\
5924\,--\,5946 & NII~5927.81,\,5931.68,\,5941.65; FeIII~5929.68\\
6558\,--\,6566 & HI~(H$\alpha$ line core)\,6562.81\\
6672\,--\,6683 & HeI~6678.15\\
\hline
\end{tabular}
\end{table*}


%%%%%%%%%%%%%%%%%%%%%%%%%%%%%%%%%%%%%%%%%%%%%%%%%%

% Don't change these lines
\bsp	% typesetting comment
\label{lastpage}
\end{document}